\documentclass[prd,aps,superscriptaddress,nofootinbib]{revtex4-2}

\usepackage{amsmath,amssymb,amsfonts,mathrsfs,braket,bm,accents}

\usepackage{graphicx}
\usepackage[colorlinks=true, linkcolor=black, citecolor=blue, urlcolor=red]{hyperref}

\begin{document}

\title{Gravitational wave interactions with matter: beyond quadrupolar perturbations}

\author{Ulrich K. {Beckering Vinckers}}
\email{ulrich.beckeringvinckers@ru.ac.za}
\affiliation{Department of Mathematics, Rhodes University, Makhanda, 6140, South Africa}
\author{Nigel T. Bishop}
\email{n.bishop@ru.ac.za}
\affiliation{Department of Mathematics, Rhodes University, Makhanda, 6140, South Africa}
\affiliation{National Institute for Theoretical and Computational Sciences (NITheCS), Stellenbosch, South Africa}

\begin{abstract}
Previous work has developed the theory of linearized gravitational wave (GW) interactions with matter using the Bondi--Sachs formalism, but with the perturbations restricted to be quadrupolar, i.e., the angular dependence is spherical harmonic with $\ell=2$. Here, the theory is extended to the case of GWs on a Minkowski background with general $\ell$. Formulas for the GW damping and heating effects are obtained for arbitrary $\ell\ge 2$. It is found that the effects are, generally, enhanced, and this suggests that it is unlikely that higher $\ell$-modes will be seen in GW observations of the post-merger signal of a binary neutron star merger, or of a core collapse supernova.
\end{abstract}

\maketitle

\section{Introduction}
Since the detection of gravitational waves (GWs)~\cite{LIGOScientific:2016aoc}, there has been recent interest in the interaction between GWs and matter~\cite{Bishop:2019ckc, Naidoo:2021rzw, Bishop:2022kzq, Kakkat:2023vni, Bishop:2024htv}. Such investigations are typically carried out by studying the linearized field equations of General Relativity (GR) through the use of the Bondi--Sachs metric~\cite{Bondi:1962px, Sachs:1962wk}, together with the so-called eth formalism~\cite{Newman:1966ub,Penrose:1985bww}. In particular, a procedure for solving the linearized field equations in such a context, and for a Minkowski background, was described in~\cite{Bishop:2004ug}. This procedure was subsequently used in~\cite{Bishop:2019ckc} to study the interaction of GWs with a spherical dust shell, and those results were applied to astrophysical scenarios in~\cite{Naidoo:2021rzw}. The damping of GWs propagating through a matter shell was studied in~\cite{Bishop:2022kzq}, with the results of such a study being used in~\cite{Kakkat:2023vni} to determine the temperature distribution inside the shell. A further study of this temperature distribution for astrophysical and cosmological scenarios was performed in~\cite{Bishop:2024oht}.

The results presented in~\cite{Bishop:2022kzq} were specific to the case where a Minkowski background is considered, and the results were extended to a Schwarzschild background in~\cite{Bishop:2024htv}. While analytical results were obtained in~\cite{Bishop:2004ug, Bishop:2022kzq} for the Minkowski case, the Schwarzschild cases studied in~\cite{Bishop:2024htv,Bishop:2025foi} required the use of numerical methods. It is worth noting that~\cite{Bishop:2019ckc, Naidoo:2021rzw, Bishop:2022kzq, Kakkat:2023vni, Bishop:2024htv, Bishop:2004ug,Bishop:2024oht,Bishop:2025foi} only considered quadrupolar contributions to the GWs.

While the studies of GW interaction with matter discussed above are specific to quadrupolar ($\ell = 2$) contributions, octupolar ($\ell = 3$) contributions to the metric components on a Minkowski background have been studied previously~\cite{Reisswig:2006nt}. In the present work, we consider a solution for the case of a Minkowski background for an arbitrary $\ell \geq 2$ mode. We also apply this solution to the consideration of GW interaction with a matter shell, which was carried out for the $\ell = 2$ case in~\cite{Bishop:2022kzq}. Such an application involves the consideration of a congruence of timelike geodesics, and analyzing the shear tensor associated with it. To this end, we find analytical expressions for the shear tensor components for a general value of $\ell$, and use these to study GW damping and heating effects. Evaluation of the general formulas in the $\ell=2$ case provides a consistency check against previous results, and evaluation in the $\ell=3$ and $4$ cases is used to consider the astrophysical implications of this work.

This manuscript is organized as follows. In Section~\ref{sec:Bondi_Sachs} we briefly review the Bondi--Sachs metric, and the eth formalism which have been used in~\cite{Bishop:2004ug} to construct a solution procedure for obtaining fixed $\ell$-mode solutions that are perturbations around a Minkowski background. We also briefly review such a solution procedure in Section~\ref{sec:Bondi_Sachs}. In Section~\ref{sec:general_ell_mode_solution}, we give the metric variables for a general $\ell \geq 2$ mode solution. In Section~\ref{sec:matter_shell}, we consider a matter shell that is treated as having no back-reaction onto the metric. There, and in Section~\ref{sec:velocity_field}, we discuss how the velocity field perturbations associated with the matter field are related to the metric perturbations through the geodesic equation as well as the normalization condition for the velocity field. We then make use of these expressions for the velocity field components to obtain expressions for the shear tensor in Section~\ref{sec:geodesic_congruence}, which is associated with a congruence of geodesics whose tangent vector field is given by the velocity field. Section~\ref{sec:damping_and_heating} examines the GW damping and heating effects in the matter shell for the $\ell = 2$, $3$, and $4$ cases. The damping effects for these different $\ell$-modes are then compared in the context of astrophysically motivated scenarios in Section~\ref{sec:astrophysical_applications}. We discuss our conclusions in Section~\ref{sec:conclusions}. We provide explicit expressions for the metric variables, velocity field components, and shear tensor components for the $\ell = 2$, $3$, and $4$ cases in the appendices. The computer algebra code used in this work is also described in the appendices, and is available as supplementary material.

\section{Previous work: the Bondi--Sachs formalism and linearized perturbations about Minkowski}\label{sec:Bondi_Sachs}
In the present work, we make use of the so-called \textit{Bondi--Sachs metric}~\cite{Bondi:1962px, Sachs:1962wk}, which is given by the following line-element
\begin{align}\label{eq:Bondi_Sachs_metric}
&\mathrm{d}s^2 = -\left[\text{e}^{2\beta}\left(1 + \frac{W}{r}\right) - r^2 h_{AB}U^AU^B\right]\mathrm{d}u^2 - 2 \text{e}^{2\beta}\mathrm{d}u \mathrm{d}r - 2 r^2 h_{AB} U^B \mathrm{d}u \mathrm{d}x^A + r^2 h_{AB} \mathrm{d}x^A \mathrm{d}x^B\,,
\end{align}
where $x^A$ are spherical polars $(\theta,\phi)$ angular coordinates. Here, and unless specified otherwise, we use lower case Latin indices to denote space-time components, and use upper case Latin indices to denote angular components. We note that $h^{AC} h_{BC} = \delta^A_B$ and impose $\text{det}(h_{AB}) = \text{det}(q_{AB})$ where $q_{A B}$ is the 2-metric of the unit 2-sphere $\mathrm{d}s^2 = \mathrm{d}\theta^2 + \sin^2\theta \mathrm{d}\phi^2$; thus $r$ is a surface area radial coordinate with each 2-sphere at constant $(u,r)$ having area $4\pi r^2$.  We make use of a complex dyad $q^A$ which satisfies $q_{AB} = q_{(A} \bar{q}_{B)}$ and is normalized to $q^A \bar{q}_A = 2$. In the present work, we use $q^A = (1, i / \sin\theta)$. We also define $U = q^A U_A$ and $J = \tfrac12 q^A q^B h_{A B}$, and use the angular ``eth'' operator, denoted by $\eth$; we follow the conventions described in Appendix 2 of ~\cite{Bishop:2016lgv}.

The Einstein field equations~\cite{Einstein:1915ca} for the case of a vacuum are simply $R_{a b} = 0$, and we linearize these around a Minkowski background. Such equations can be re-cast~\cite{Bishop:2004ug} as seven equations by performing projections of the Ricci tensor with the complex dyad $q^A$. More specifically, these equations are given by $R_{11}$, $q^A R_{A 1}$, $q^A q^B R_{A B}$, $h^{A B} R_{A B}$, $R_{0 0}$, $R_{0 1}$, and $q^A R_{A 0}$, each set equal to zero. In Eqs.~\eqref{eq:R11}--\eqref{eq:qA_RA0} of Appendix~\ref{sec:linearised_field_equations}, we give these seven linearized expressions for the case of a vacuum, which were previously reported in~\cite{Bishop:2004ug}.

One can find solutions to the linearized field equations by making use of the following ansatz:
\begin{align}\label{eq:metric_ansatz}
\beta &= \Re\left({\text e}^{i \nu u} \beta^{[\ell, \ell]}(r)\right) \,_0 Z_{\ell, \ell} \,, \hspace{1.7cm} J = \Re\left({\text e}^{i \nu u} J^{[\ell, \ell]}(r)\right) \,_2 Z_{\ell, \ell} \,, \nonumber \\
U &= \Re\left({\text e}^{i \nu u} U^{[\ell, \ell]}(r)\right) \,_1 Z_{\ell, \ell} \,, \hspace{1.45cm} W = \Re\left({\text e}^{i \nu u} W^{[\ell, \ell]}(r)\right) \,_0 Z_{\ell, \ell} \,,
\end{align}
which has been made use of previously for the $\ell = 2$~\cite{Bishop:2019ckc} and the $\ell = 3$~\cite{Reisswig:2006nt} cases\footnote{We note that the ansatz used in~\cite{Reisswig:2006nt} differs to the one used here by a factor of $\sqrt{(\ell + 2) (\ell + 1) \ell (\ell - 1)}$ in the ansatz for $J$, and a factor of $\sqrt{\ell (\ell + 1)}$ in the ansatz for $U$.}. Here, the $\,_s Z_{\ell, \ell}$ are angular basis functions which are related (see for example Eq.~(428) of~\cite{Bishop:2016lgv}) to angular derivatives of the spherical harmonics $Y_{\ell, m}$. A solution procedure for a fixed value of $\ell$ has been outlined previously in~\cite{Bishop:2004ug}, and discussed further in~\cite{Bishop:2019ckc}. We now wish to briefly describe this procedure, which is applied after substituting the ansatz given in Eq.~\eqref{eq:metric_ansatz} into the field equations. 

Firstly, one solves the $R_{11}$ equation which results in $\beta^{[\ell, \ell]}$ being a constant. Subsequently, the $q^A R_{A1}$ and $q^A q^B R_{A B}$ equations are solved together to obtain expressions for $U^{[\ell, \ell]}$ and $J^{[\ell, \ell]}$. Upon obtaining these expressions, one can integrate the $h^{AB} R_{AB}$ equation to find $W^{[\ell, \ell]}$. Two constants of integration are then fixed by the $R_{00}$ equation, and one is left to verify that the $R_{01}$ and $q^A R_{A 0}$ equations are satisfied.

In the following section, we consider a general single $\ell$-mode solution to the vacuum field equations linearized around a Minkowski background.

\section{Metric for a general $\ell$-mode solution}\label{sec:general_ell_mode_solution}

We start by considering the $R_{11}$, $q^A R_{A1}$, and $q^A q^B R_{A B}$ equations. In order for the $R_{11}$ equation to be satisfied, we require that $\beta$ be independent of $r$, i.e.,
\begin{align}\label{eq:general_beta_solution}
\beta^{[\ell, \ell]} &= b_0 \,.
\end{align}
Having found the solution for $\beta$, the $q^A R_{A1}$ and $q^A q^B R_{A B}$ equations now provide us with differential equations for $U$ and $J$. Imposing the physical condition that there are no incoming GWs, we find that these two equations are solved when 
\begin{align}
U^{[\ell, \ell]} &= U_\infty + \frac{2 \sqrt{\ell (\ell + 1)} (b_0r + C_{30})}{r^2} - \frac{8 i \nu \sqrt{\ell (\ell + 1)} C_{40}}{\sqrt{(\ell + 2) (\ell - 1)}} \sum_{n=3}^{\ell + 2} r^{-n} (n - 1) \Omega_{n, \ell} \,, \label{eq:general_U_solution} \\
J^{[\ell, \ell]} &= \frac{i \sqrt{(\ell + 2) (\ell - 1)} U_\infty}{\nu} + \frac{4 \sqrt{\ell (\ell + 1)}}{\sqrt{(\ell + 2) (\ell - 1)}}\left(\frac{C_{30}}{r} + \frac{2 i \nu C_{40}}{\sqrt{(\ell + 2) (\ell - 1)}} \sum_{n=3}^{\ell + 2} r^{-(n - 1)} n (n - 3) \Omega_{n, \ell} \right)\,, \label{eq:general_J_solution}
\end{align}
where $\ell \geq 2$ and the $\Omega_{n, \ell}$ coefficients are given by
\begin{align}\label{eq:Omega_plus_1}
\Omega_{3, \ell} = 1 / 2 \,, \hspace{1.5cm}
\Omega_{n + 1, \ell} = \Omega_{n, \ell} \frac{\left[(\ell + 2) (\ell - 1) - n (n - 3)\right]}{2 (n + 1) i \nu} \,,
\end{align}
where $U_\infty$, $C_{30}$, and $C_{40}$ are constants. One can verify that Eqs.~\eqref{eq:general_beta_solution}--\eqref{eq:general_J_solution} satisfy the $R_{11}$, $q^A R_{A1}$, and $q^A q^B R_{A B}$ equations through direct substitution. In performing such a verification, it is useful to note that $\Omega_{n, \ell} = 0$ for $n \geq \ell + 3$, which is easily seen from Eq.~\eqref{eq:Omega_plus_1}. Therefore, one can write the summations over $n$ in the expressions for the metric variables as infinite sums. Nevertheless, we have left them as summations up to $n = \ell + 2$ in order to explicitly indicate the highest order in $1 / r$ that appears for each metric variable. 

We now turn our attention to the $h^{AB} R_{AB}$ equation. After substituting in the ansatz Eq.~\eqref{eq:metric_ansatz}, one can write the $h^{AB} R_{AB}$ equation in terms of the radial parts of the the metric variables. Substituting Eqs.~\eqref{eq:general_beta_solution}--\eqref{eq:general_J_solution} into the resulting expression and integrating yields the following for $W^{[\ell, \ell]}$
\begin{align}\label{eq:general_W_solution}
W^{[\ell, \ell]} = C_W + 2 \left[1 - \ell (\ell + 1) \right] b_0 r - r \sqrt{\ell (\ell + 1)} U_{\infty}\left(r - \frac{i (\ell + 2) (\ell - 1)}{2 \nu}\right) - \frac{8 i \nu \ell (\ell + 1) C_{40}}{\sqrt{(\ell + 2) (\ell - 1)}} \sum_{n = 3}^{\ell + 2} r^{-(n - 2)} \Omega_{n, \ell} \,,
\end{align}
where $C_W$ is a constant of integration.

We are now left with the $R_{00}$, $R_{01}$, and $q^A R_{A 0}$ equations. Using the expressions given above for the metric variables in the $R_{00}$ equation, one obtains constraints for the constants of integration $C_W$ and $C_{30}$. We find:
\begin{align}
& \hspace{5cm} C_W = \frac{24 \nu^2 C_{40}}{\sqrt{(\ell + 2) (\ell - 1)}} \,, \label{eq:C_constraint} \\
&C_{30} = \frac{\left[\ell (\ell + 1) (\ell + 2) (\ell - 1)\right]^{3 /2} U_\infty - 96 \nu^4 C_{40} +  4 i \nu \ell (\ell + 1) (\ell + 2)^{3 / 2} (\ell - 1)^{3 / 2} b_0}{8 \nu^2 \ell (\ell + 1) \sqrt{(\ell + 2) (\ell - 1)}} \label{eq:U_infty_constraint} \,.
\end{align}
The $R_{01}$ equation is satisfied regardless of whether the constraint Eqs.~\eqref{eq:C_constraint} and~\eqref{eq:U_infty_constraint} are imposed. On the other hand, it is necessary to impose the constraint Eq.~\eqref{eq:C_constraint} in order for the $q^A R_{A0}$ equation to be satisfied. It will be found (see Eq.
\eqref{eq:News} below) that $C_{40}$ is physical and is fixed by the power output of GWs, and the remaining constants, $b_0$ and $U_\infty$, are gauge freedoms whose values can be freely chosen. We will not set values for them since keeping them free provides a consistency check: any physical quantity should be gauge independent and so depend on $C_{40}$ only. However, it is worth noting that in some applications it is useful to work in a gauge that is explicitly Minkowskian as $r\rightarrow\infty$ (known as the Bondi gauge), which is achieved by setting $b_0=U_\infty=0$.

In Appendix~\ref{sec:specific_ell_metric}, we give the metric variables for the $\ell=2$, $3$, and $4$ cases, and compare the $\ell=2$ and $3$ expressions to those given in~\cite{Bishop:2022kzq} and~\cite{Reisswig:2006nt}, respectively.

\section{Matter shell}\label{sec:matter_shell}
\subsection{Velocity field}\label{sec:velocity_field}
Having obtained expressions for the metric variables, we now turn our attention to a matter shell which we treat as having no back-reaction onto the metric, i.e., it is a test shell and has a negligible contribution\footnote{For a discussion on the validity of such an approximation, we direct the reader to Appendix~\ref{sec:validity_discussion}.} to the field equations. In addition, we take this shell to be viscous, with a coefficient of shear viscosity $\eta$. The timelike four-velocity is denoted by $V^a$ and it is normalized to unity, i.e., $V^a V_a = -1$. The background four-velocity is simply $(\partial_u)^a$, and it satisfies the normalization condition using the Minkowski metric in Bondi--Sachs form. Expressions for $V_1$ and $V_A$ can be obtained through the use of the conservation equation which in this case gives the geodesic equation: $V^b \nabla_b V^a = 0$. From this geodesic equation, one can obtain the following for $V_1$ and $V_{\text{ang}} = q^A V_A$
\begin{align}\label{eq:V1_ang_eths}
\partial_u (V_1, V_{\text{ang}}) = - (\partial_r, \eth) \left(\beta + \frac{W}{2 r}\right) \,.
\end{align}
On the other hand, one can obtain an expression for $V_0$ in terms of the metric components through the use of the normalization condition. Such a condition gives
\begin{align}\label{eq:V0_eths}
V_0  = -1 - \beta - \frac{W}{2 r} \,.
\end{align}
Eqs.~\eqref{eq:V1_ang_eths} and~\eqref{eq:V0_eths} allow us to obtain expressions for the velocity field components given the metric variables. We now wish to make use of these to study a congruence of curves whose tangent vector is given by the velocity field. In such a context, we shall examine the shear associated with the congruence, which will be used in subsequent sections to study GW damping and heating.

\subsection{Congruence of timelike geodesics}\label{sec:geodesic_congruence}
Let us now consider the type $(0, 2)$ tensor field $B_{ab} := \nabla_b V_a$ which can be projected into a hypersurface with normal vector field $V^a$ through $P_a\,^c P_b\,^d B_{cd}$ where $P_a\,^b := \delta_a^b + V_a V^b$ is the projection operator. This projected tensor field can be decomposed~\cite{Wald:1984rg} to give the vorticity, shear, and expansion for a congruence of curves whose tangent vector field is $V^a$. More specifically, the antisymmetric part gives the \textit{vorticity tensor} $w_{ab} $, the trace-free symmetric part is the \textit{shear tensor} $\sigma_{ab}$, and the trace is the \textit{expansion} $\Theta$. However, since $V^a$ satisfies the geodesic equation, which gives $V^b B_{ab} = 0$, while also being normalized to unity, which gives $V^a B_{ab} = 0$, we simply have $P_a\,^c P_b\,^d B_{cd} = B_{ab}$.

In the present work, we shall study GW damping by making use of the energy loss per unit volume through the shell which is given by $-2 \eta \sigma_{ab} \sigma^{ab}$~\cite{Baumgarte:2010ndz}. Noting that $B_{ab}$ comprises only linear order terms, and that the only zeroth order component of $V^a$ is $V^0=1$, the conditions $V^b B_{ab} = 0$ and $V^b B_{ab} = 0$ imply $B_{(0a)} = 0$ to linear order.
For the remaining components, we find
\begin{align}
B_{11} &= \partial_r V_1 + 2 \partial_r \beta\,, \label{eq:B_11}\\ 
q^A q^B B_{AB} &= r^2 \eth U + r^2 \partial_u J + \eth V_{\text{ang}} \,, \\
2 q^A B_{(1A)} &= r^2 \partial_r U + \eth V_1 + 2\eth \beta + \partial_r V_{\text{ang}} - \tfrac{2}{r} V_{\text{ang}} \,, \label{eq:B_1A} \\
q^A \bar{q}^B B_{(AB)} &= \tfrac{r^2}{2} \left(\bar{\eth} U + \eth \bar{U}\right) + 2 r \left(V_1 - V_0\right) - 2 W + \tfrac12 \left(\bar{\eth} V_{\text{ang}} + \eth \bar{V}_{\text{ang}}\right) \,. \label{eq:B_W}
\end{align}
We find that the expansion associated with the congruence is given by
\begin{align}\label{eq:Theta}
\Theta = \tfrac12\left(\bar{\eth}U + \eth\bar{U}\right) + \left(\partial_r + \tfrac2r\right) (V_1 - V_0) - \partial_u V_1 - \frac{\partial_r (rW)}{r^2} + \frac{\bar{\eth} V_{\text{ang}} + \eth \bar{V}_{\text{ang}}}{2 r^{2}} \,.
\end{align}
At this point, we have not assumed a specific solution for the metric variables in the expressions for the $B_{ab}$ tensor field. The expansion of GWs is zero\footnote{The vanishing of the expansion for the solutions considered here can be verified as follows. By taking the derivative of Eq.~\eqref{eq:Theta} with respect to $u$, and substituting the velocity field components for metric variables through the use of Eqs.~\eqref{eq:V1_ang_eths} and~\eqref{eq:V0_eths}, one can show that $\partial_u \Theta$ is nothing more than the left-hand side of the $R_{00}$ Eq.~\eqref{eq:R00} with a minus sign, and is therefore zero. On the other hand, substituting the solution given in Section~\ref{sec:general_ell_mode_solution} into Eq.~\eqref{eq:Theta} results in the expansion being of the form $\Theta = \Re(\Theta^{[\ell, \ell]}(r) \mathrm{e}^{i \nu u}) \,_0 Z_{\ell, \ell}$. It follows then that $\partial_u \Theta = 0$ gives $\Theta = 0$.}, so we have $q^A \bar{q}^B B_{(AB)} = -r^2 B_{11}$. Since  we are considering a congruence of normalized timelike geodesics, we can compute the shear tensor through $\sigma_{ab} = B_{(ab)}$. We now compute $\sigma_{ab} \sigma^{ab}$ through
\begin{align}\label{eq:sigma_squared}
\sigma_{ab} \sigma^{ab} = \tfrac32 \sigma_{11}^2 + \tfrac2{r^2} |q^A \sigma_{1A}|^2 + \tfrac1{2 r^4}|q^A q^B \sigma_{AB}|^2 \,.
\end{align}
For a derivation of Eq.~\eqref{eq:sigma_squared}, we direct the reader to Appendix~\ref{sec:derivation_of_sigma_squared}. 

Later calculations of the damping and temperature effects will involve this quantity integrated over the sphere and time-averaged\footnote{We note that we compute the time average of a function, say $f$, through 
%\begin{equation}\label{eq:time_average_def}
$\langle f \rangle = \frac{\nu}{2 \pi} \int_0^{\frac{2\pi}{\nu}} \mathrm{d}u f \,.
$%\end{equation}
} over a wave period of $2 \pi / \nu$. More specifically, and as will be discussed further in Section~\ref{sec:damping_and_heating}, we use the energy loss through the shell to derive the GW damping factors, and are thus required to integrate $\sigma_{ab} \sigma^{ab}$ over the shell. In addition, the energy loss per unit volume appears as a source term in the diffusion equation which is solved to find the temperature increase through the shell. In obtaining such a temperature increase, we will assume that the heating effect is uniform, and will therefore average over the angular directions by integrating over $\theta$ and $\phi$. We therefore compute
\begin{align}\label{eq:integrated_sigma_squared}
\left\langle\int \mathrm{d}\Omega \ \sigma_{ab} \sigma^{ab} \right\rangle = \tfrac34 \left|\sigma^{[\ell, \ell]}_{11}\right|^2 + \left|\sigma^{[\ell, \ell]}_{U} / r\right|^2 + \tfrac14 \left|\sigma^{[\ell, \ell]}_J / r^2\right|^2 \,,
\end{align}
where we have used
\begin{equation}\label{eq:shear:ansatz}
\sigma_{11} = \Re\left(\sigma_{11}^{[\ell, \ell]}(r) \text{e}^{i u \nu}\right) \,_0Z_{\ell, \ell} \,, \hspace{1cm} q^A \sigma_{1A} = \Re\left(\sigma_{U}^{[\ell, \ell]}(r) \text{e}^{i u \nu}\right) \,_1Z_{\ell, \ell} \,, \hspace{1cm} q^A q^B \sigma_{AB} = \Re\left(\sigma_{J}^{[\ell, \ell]}(r) \text{e}^{i u \nu}\right) \,_2Z_{\ell, \ell} \,.
\end{equation}
We note that a similar decomposition of the shear tensor was used for the $\ell = 2$ case in~\cite{Bishop:2022kzq}. 

We now wish obtain explicit expressions for the radial parts of the shear tensor components given the general $\ell$-mode solution discussed in the previous section. To this end, one first needs to obtain expressions for the velocity field, $V_a$. This is done by using an ansatz for the velocity field components which is similar\footnote{See Eq.~\eqref{eq:velocity_fields_ansatz} of Appendix~\ref{sec:specific_ell_metric} for the explicit ansatz used for the velocity field components. It is important to note that such an ansatz is used for the components of the dual field, i.e., $V_a$.} to the ones used for the metric variables, and then making use of Eqs.~\eqref{eq:V1_ang_eths} and~\eqref{eq:V0_eths}. Upon obtaining expressions for the velocity field perturbations, one can then make use of Eqs.~\eqref{eq:B_11}--\eqref{eq:B_1A} to find expressions for the shear tensor components; recall that $q^A \bar{q}^B \sigma_{AB} = -r^2 \sigma_{11}$. For the linearized solution described in Section~\ref{sec:general_ell_mode_solution}, we find the following for the shear tensor components
\begin{align}
\sigma^{[\ell, \ell]}_{11} &= \frac{4 \ell (\ell + 1) C_{40}}{\sqrt{(\ell + 2) (\ell - 1)}} \sum_{n=0}^{\ell + 2} r^{-(n + 1)} \Omega_{n, \ell} n (n - 1) \,, \label{eq:sigma_11_ell}\\
\sigma^{[\ell, \ell]}_J / r^2 &= \frac{2\sqrt{\ell (\ell + 1)} C_{40}}{(\ell + 2) (\ell - 1)} \sum_{n = 0}^{\ell + 2} r^{-(n + 1)} \Omega_{n, \ell} \left[(\ell + 2) (\ell + 1) \ell (\ell - 1) + n (n - 1) (n - 2) (n - 3)\right] \,, \label{eq:sigma_J_ell}\\
\sigma^{[\ell, \ell]}_{U} / r &= -\frac{2 \sqrt{\ell (\ell + 1)} C_{40}}{\sqrt{(\ell + 2) (\ell - 1)}}\sum_{n=0}^{\ell + 2} r^{-(n + 1)} \Omega_{n, \ell} n \left[\ell (\ell + 1) + (n - 1) (n - 2)\right] \,. \label{eq:sigma_U_ell}
\end{align}
Eqs.~\eqref{eq:sigma_11_ell}--\eqref{eq:sigma_U_ell} given above provide us with expressions for the shear tensor for an arbitrary $\ell$-mode for $\ell \geq 2$. In Appendix~\ref{sec:specific_ell_metric}, we make use of these expressions to write the shear tensor components for the $\ell = 2$, $3$, and $4$ cases.

\subsection{GW damping and temperature increase}\label{sec:damping_and_heating}
Having determined the shear tensor components for a general $\ell$-mode, we now wish to use these to compute the rate of energy loss $\langle{\dot E_{\text{shell}}}\rangle$ through the matter shell. This is obtained by integrating $-2 \eta \sigma_{ab} \sigma^{ab}$, which gives the rate of energy loss per unit volume, over a shell of radius $r$ and thickness $\delta r$, and averaging over a wave period:
\begin{equation}\label{eq:E_shell_integral}
\langle{\dot E_{\text{shell}}}\rangle = -2 \eta \left\langle \int_{r}^{r + \delta r} \mathrm{d}r' r'^2 \int \mathrm{d}\Omega \ \sigma_{ab} \sigma^{ab} \right\rangle \,,
\end{equation}
where $\sigma_{ab} = \sigma_{ab}(r', \theta, \phi)$ in the integrand. Thus, up to first order in the shell thickness $\delta r$, $\langle{\dot E_{\text{shell}}}\rangle$ is given by Eq.~\eqref{eq:integrated_sigma_squared} multiplied by a factor of $-2 \eta \delta r r^2$.

Following~\cite{Bishop:2022kzq} where the $\ell = 2$ case is considered, we wish to write the energy loss per unit volume in terms of the rate of energy $\langle\dot{E}_{\text{GW}}\rangle$ that is being output as GWs. This latter quantity can be obtained by first computing $\tfrac{1}{4\pi} \int \mathrm{d}\Omega| N_{\ell, \ell}|^2$, where $N_{\ell, \ell}$ is the \textit{Bondi news}~\cite{Bondi:1962px}, and then averaging over a wave period:
\begin{equation}\label{eq:E_GW_from_N}
\langle \dot{E}_{\text{GW}} \rangle = \left\langle \frac{1}{4\pi} \int \mathrm{d}\Omega| N_{\ell, \ell}|^2 \right\rangle \,.
\end{equation}
Here, we compute the news by making use\footnote{Note that the $J_0$ used in~\cite{Bishop:2004ug} is related to $J^{[\ell, \ell]}$ used here through $J^{[\ell, \ell]} = \sqrt{(\ell + 2) (\ell + 1) \ell (\ell - 1)} J_0$. In addition, we take the real part of the factor involving the $u$ and $r$ dependence; this allows for $\,_2Z_{\ell, \ell}$ to be the only complex part in the news. This is similar to what is done in~\cite{Reisswig:2006nt}.} of Eq.~(31) of~\cite{Bishop:2004ug}. To this end, we find
\begin{equation}
N_{\ell, \ell} = \frac{24 \nu^3 \Re\left(-iC_{40} {\text e}^{i \nu u}\right)}{(\ell + 2) (\ell - 1) \sqrt{\ell (\ell + 1)}} \,_2 Z_{\ell, \ell}\,.
\label{eq:News}
\end{equation}
In Appendix~\ref{sec:specific_ell_metric}, we evaluate Eq.~\eqref{eq:News} to obtain the news for the cases where $\ell=2$, $3$, and $4$. We note that $N_{2, 2}$ agrees with the news given in the literature (see for example~\cite{Bishop:2022kzq}). In addition, we find that $N_{3, 3}$ agrees\footnote{See Appendix~\ref{sec:ell_3_expressions} of the present work for a comparison between the notation used here and the notation used in~\cite{Reisswig:2006nt}.} with the news given in the second equality of Eq.~(16) in~\cite{Reisswig:2006nt}.

We now find the following for the rate of energy that is being output as GWs by substituting Eq.~\eqref{eq:News} into Eq.~\eqref{eq:E_GW_from_N} and obtain:
\begin{equation}\label{eq:E_GW}
\langle{\dot{E}_{\text{GW}}}\rangle = \frac{72 \nu^6 |C_{40}|^2}{\pi (\ell + 2)^2 (\ell - 1)^2 \ell (\ell + 1)} \,.
\end{equation}
We have also evaluated Eq.~\eqref{eq:E_GW} for the same specific values of $\ell$ mentioned above, and provided these in Appendix~\ref{sec:specific_ell_metric}. We note that the $\ell=2$ expression for $\langle{\dot{E}_{\text{GW}}}\rangle$ has been reported previously in Eq.~(13) of~\cite{Bishop:2022kzq}. 

In order for energy to be conserved through the matter shell, we have
\begin{equation}
\langle \dot{E}_{\text{GW}}\rangle (r + \delta r) = \langle \dot{E}_{\text{GW}}\rangle (r) + \langle \dot{E}_{\text{shell}}\rangle (r) \,.
\end{equation}
We now use $\langle\dot{E}_{\text{GW}}\rangle \propto H^2$, where $H$ is the rescaled GW magnitude, in order to write
\begin{equation}
H^2(r+\delta r) = H^2(r)\left(1 + \frac{\langle \dot{E}_{\text{shell}}\rangle}{\langle \dot{E}_{\text{GW}}\rangle}\right) \,,
\end{equation}
which, after taking the square root and Taylor expanding, gives the following
\begin{align}\label{eq:H_r_+_delta_r}
H(r + \delta r) \approx H(r) \left(1 + \frac{\langle\dot{E}_{\text{shell}}\rangle}{2 \langle\dot{E}_{\text{GW}}\rangle}\right) \hspace{1cm} \implies \hspace{1cm} \frac{\mathrm{d} \ln H}{\mathrm{d} r} \approx \frac{\langle\dot{E}_{\text{shell}}\rangle}{2  \delta r \langle\dot{E}_{\text{GW}}\rangle} \,.
\end{align}
We can now integrate the last expression in order to obtain the rescaled GW magnitude up to zeroth order in $\delta r$
\begin{align}\label{eq:H_ell}
H = \text{exp}\left[-\frac{\eta}{\langle{\dot{E}_{\text{GW}}}\rangle} \int \mathrm{d}r r^2 \left\langle\int \mathrm{d}\Omega \ \sigma_{ab} \sigma^{ab} \right\rangle\right] \,,
\end{align}
and where we have used Eq.~\eqref{eq:E_shell_integral} for $\langle \dot{E}_{\text{shell}}\rangle$. The process for evaluating Eq.~\eqref{eq:H_ell} is as follows. We substitute the shear tensor components given in Eqs.~\eqref{eq:sigma_11_ell}--\eqref{eq:sigma_U_ell} into the time averaged expression in Eq.~\eqref{eq:integrated_sigma_squared}. The resulting expression along with Eq.~\eqref{eq:E_GW} is then substituted into Eq.~\eqref{eq:H_ell} to obtain an expression for the rescaled GW magnitude. It is worth noting that the $|C_{40}|^2$ factors in $\langle\dot{E}_{\text{GW}}\rangle$ cancel with those contained in the expressions for the shear tensor components. Below, we evaluate $H$ for the $\ell = 2$, $3$, and $4$ cases explicitly:
\begin{align}
H_{\ell=2} &= C_{\ell=2} \exp\left[- 8 \pi \eta \left(r - \frac{2}{\nu_{\ell=2}^{2} r} - \frac{3}{\nu_{\ell=2}^{4} r^{3}} - \frac{9}{\nu_{\ell=2}^{6} r^{5}} - \frac{45}{\nu_{\ell=2}^{8} r^{7}}\right)\right] \,, \label{eq:H_ell_2}\\
H_{\ell=3} &= C_{\ell=3} \exp\left[- 8 \pi \eta \left(r - \frac{5}{\nu_{\ell=3}^{2} r} - \frac{15}{\nu_{\ell=3}^{4} r^{3}} - \frac{90}{\nu_{\ell=3}^{6} r^{5}} - \frac{675}{\nu_{\ell=3}^{8} r^{7}} - \frac{4725}{\nu_{\ell=3}^{10} r^{9}}\right)\right] \,, \label{eq:H_ell_3} \\
H_{\ell=4} &= C_{\ell=4} \exp\left[- 8 \pi \eta \left(r - \frac{9}{\nu_{\ell=4}^{2} r} - \frac{45}{\nu_{\ell=4}^{4} r^{3}} - \frac{450}{\nu_{\ell=4}^{6} r^{5}} - \frac{5400}{\nu_{\ell=4}^{8} r^{7}} - \frac{66150}{\nu_{\ell=4}^{10} r^{9}} - \frac{595350}{\nu_{\ell=4}^{12} r^{11}}\right)\right] \,, \label{eq:H_ell_4}
\end{align}
where the $C_\ell$ are constants of integration, which can be fixed if the $H_\ell$ are known at a given value of $r$. We note that Eq.~\eqref{eq:H_ell_2} has been reported previously in~\cite{Bishop:2022kzq}.

We can also compute the temperature increase through the shell. Such an analysis for the $\ell = 2$ case has been performed in~\cite{Kakkat:2023vni, Bishop:2024oht}. In~\cite{Kakkat:2023vni}, it is discussed how one can obtain the temperature increase by solving the diffusion equation with a source term, $f$, which is related to the shear through $f = 2 \eta \langle \sigma_{ab} \sigma^{ab}\rangle / (C \rho)$, where $\rho$ and $C$ are, respectively, the density and specific heat capacity. In the $\ell = 2$ case, solving the diffusion equation gives a solution that is a sum of $Y_{0, 0}$, $Y_{2, 0}$, and $Y_{4, 0}$ terms. By assuming that the heating effect is uniform, the angular dependent parts are discarded~\cite{Bishop:2024oht} and one is left with only the $Y_{0, 0}$ part. The same result may be obtained by averaging over the sphere. Thus, we find the following expression for the temperature increase
\begin{align}
\Delta T = \frac{u \eta (\ell + 2)^2 (\ell - 1)^2 \ell (\ell + 1) \langle{\dot{E}_{\text{GW}}}\rangle}{144 C \rho \nu^6 |C_{40}|^2} \left\langle\int \mathrm{d}\Omega \ \sigma_{ab} \sigma^{ab} \right\rangle \,,
\end{align}
where $\Delta T = T - T_0$ and $T_0$ is the temperature at $u=0$.

Below, we write the temperature increase for the $\ell = 2$, $3$, and $4$ cases:
\begin{align}
\Delta T_{\ell = 2} &= \frac{4 \eta \langle\dot{E}_{\text{GW}}\rangle_{\ell = 2} u \left(\nu_{\ell=2}^{8} r^{8} + 2 \nu_{\ell=2}^{6} r^{6} + 9 \nu_{\ell=2}^{4} r^{4} + 45 \nu_{\ell=2}^{2} r^{2} + 315\right)}{C \nu_{\ell=2}^{8} \rho r^{10}} \,, \\
\Delta T_{\ell = 3} &= \frac{4 \eta \langle\dot{E}_{\text{GW}}\rangle_{\ell = 3} u \left(\nu_{\ell=3}^{10} r^{10} + 5 \nu_{\ell=3}^{8} r^{8} + 45 \nu_{\ell=3}^{6} r^{6} + 450 \nu_{\ell=3}^{4} r^{4} + 4725 \nu_{\ell=3}^{2} r^{2} + 42525\right)}{C \nu_{\ell=3}^{10} \rho r^{12}} \,, \\
\Delta T_{\ell = 4} &= \frac{4 \eta \langle\dot{E}_{\text{GW}}\rangle_{\ell = 4} u \left(\nu_{\ell=4}^{12} r^{12} + 9 \nu_{\ell=4}^{10} r^{10} + 135 \nu_{\ell=4}^{8} r^{8} + 2250 \nu_{\ell=4}^{6} r^{6} + 37800 \nu_{\ell=4}^{4} r^{4} + 595350 \nu_{\ell=4}^{2} r^{2} + 6548850\right)}{C \nu_{\ell=4}^{12} \rho r^{14}} \,.
\end{align}

\section{Astrophysical applications}\label{sec:astrophysical_applications}
In this section, we wish to compare the rescaled GW magnitude for different values of $\ell$, and the units will be SI rather than geometric used in the previous sections. For such an investigation, it is useful to write Eqs.~\eqref{eq:H_ell_2}--\eqref{eq:H_ell_4} as
\begin{align}
H_{\ell}(r_o) =& H_{\ell}(r_i) \exp\left(-\frac{8 \pi \eta r_i G D_\ell}{c^3}\right)\,,\;\;\mbox{where}\nonumber\\
D_2=&(\alpha-1)+2 \psi_{\ell=2}^{2}(1-\alpha^{-1})+3\psi_{\ell=2}^{4}(1-\alpha^{-3})
+9 \psi_{\ell=2}^{6}(1-\alpha^{-5})+ 45 \psi_{\ell=2}^{8}(1-\alpha^{-7})\,,\nonumber\\
D_3=&(\alpha-1)+5 \psi_{\ell=3}^{2}(1-\alpha^{-1})+15\psi_{\ell=3}^{4}(1-\alpha^{-3})
+90 \psi_{\ell=3}^{6}(1-\alpha^{-5})+ 675 \psi_{\ell=3}^{8}(1-\alpha^{-7})
+ 4725 \psi_{\ell=3}^{10}(1-\alpha^{-9})\,,\nonumber\\
D_4=&(\alpha-1)+9 \psi_{\ell=4}^{2}(1-\alpha^{-1})+45\psi_{\ell=4}^{4}(1-\alpha^{-3})
+450 \psi_{\ell=4}^{6}(1-\alpha^{-5})+ 5400 \psi_{\ell=4}^{8}(1-\alpha^{-7})\nonumber\\
&+66150\psi_{\ell=4}^{10}(1-\alpha^{-9})+ 595350\psi_{\ell=4}^{12}(1-\alpha^{-11})\,,
\label{e-HD}
\end{align}
where $r_i,r_o$ are the inner and outer radii of the matter shell and $\alpha = r_o / r_i$, and where $\psi_{\ell} = c/(\nu_\ell r_i)=\lambda_{\ell} / (2 \pi r_i)$ with $\lambda_\ell$ being the wavelength. We note that $H_{\ell=2}(r_o)/H_{\ell=2}(r_i)\approx 1$ means that there is no GW damping, whereas $H_{\ell=2}(r_o)/H_{\ell=2}(r_i)\approx 0$ means that the GWs are completely damped.

In order to compare the different $H$ values, we need to establish relationships between the different wavelengths, and therefore the different $\psi_{\ell}$ values. Here, we will look at astrophysically motivated scenarios to obtain such relationships. Astrophysical applications of the GW damping and heating effects for the dominant $\ell=2$ mode have been discussed in previous work: specifically the post merger GWs from a binary neutron star (BNS) merger, GWs from a core collapse supernova (CCSNe), and the effect of matter accretion at a binary black hole (BBH) merger. These effects have been investigated on both a Minkowski background~\cite{Bishop:2022kzq,Kakkat:2023vni,Bishop:2024oht} as well as on a general spherically symmetric background~\cite{Bishop:2025foi}. Importantly, it was found that using the more realistic general spherically symmetric background  could lead to GW damping/heating effects that are larger by a factor ${\mathcal O}(10)$, but the solutions are not expressible in terms of elementary functions and have to be obtained numerically.

Both the magnitude and frequency $f=\nu/(2\pi)$ of a ${}_2Z_{\ell,\ell}$ component of GWs depend strongly on $\ell$, but the dependence varies according to the GW source; importantly, the damping/heating effects are highly sensitive to the frequency. In the following, we consider four scenarios for the GW source for different values of $\ell$. For each of these, we examine the dependence of the frequency on $\ell$. We shall also make use of the dimensionless quantities $b_{\ell = 3} = f_{\ell = 3} / f_{\ell = 2}$ and $b_{\ell = 4} = f_{\ell = 4} / f_{\ell = 2}$, and note that $b_\ell=\psi_{\ell=2}/\psi_\ell$.
\begin{itemize}
\item \textit{Masses in circular orbit:} In the linearized regime, the frequency behaves as $f\propto\ell$.
\item \textit{Schwarzschild black hole:} QNM frequencies for a Schwarzschild BH have been obtained in~\cite{Chandrasekhar:1975zza} through the use of the so-called Zerilli equation~\cite{Zerilli:1970wzz}. The Schwarzschild QNM frequencies are also reviewed in Table 1 of~\cite{Kokkotas:1999bd} for the $\ell = 2$, $3$, and $4$ cases, and using the fundamental ($n=0$) mode values we find
\begin{equation}
b_{\ell = 3} \approx 1.60 \,, \hspace{1.5cm} b_{\ell = 4} \approx 2.17 \,.
\end{equation}
\item \textit{Binary black holes:} Table 1 of~\cite{Kamaretsos:2011um} provides dimensionless $\nu M$ ringdown frequencies for the merger of a pair of nonspinning black holes; obtained through numerical relativity simulations. Here, let us consider these values when the mass ratio is $q = 2$, $3$, $4$, or $11$, which have corresponding BH spin values of $j = 0.62$, $0.54$, $0.47$ and $0.25$, respectivly. Using the values given in Table 1 of~\cite{Kamaretsos:2011um}, we find
\begin{equation}
1.57 \lesssim b_{\ell = 3} \lesssim 1.64 \,, \hspace{2cm} 2.14 \lesssim b_{\ell = 4} \lesssim 2.18 \,.
\label{e-b_BBH}
\end{equation}
In addition,  Fig. 2 of~\cite{Kamaretsos:2011um} shows how the frequencies of the different $\ell$-modes evolve during a numerical simulation of the late inspiral, merger, and ringdown of a BBH system. Although numerical values are not tabulated, it is clear from the Figure that Eq.~\eqref{e-b_BBH} is at least approximately satisfied throughout this period. Also, Fig. 3 of~\cite{Kamaretsos:2011um} shows the relative magnitudes of the different $\ell$-modes at various times during the numerical evolution and for various values of $q$. For $q\le 4$, $H_{\ell=4}(r_i)/H_{\ell=2}(r_i)<0.1$.
\item \textit{Hypermassive neutron star:} For this case, we shall approximate the fundamental mode\footnote{We note that we only consider fundamental ($n=0$) modes in this case. For a discussion of pressure and gravity modes in the context of non-rotating stars, we direct the interested reader to~\cite{Cowling:1941nqk}. For a discussion of rotational modes, see~\cite{Kokkotas:1999bd}.} frequencies using those corresponding to an irrotational Newtonian star of uniform density, and a discussion of the relevance of such an approximation for the fundamental mode of a neutron star can be found in~\cite{Kokkotas:1999bd}. An analytical expression for such frequencies has been derived in~\cite{Kelvin:1890} (see pp.~384--386), and it is given in Eq. (57) of~\cite{Kokkotas:1999bd}. Using this equation, we find
\begin{equation}\label{eq:HNS_frequencies}
b_{\ell = 3} \approx 1.46 \,, \hspace{1.5cm} b_{\ell = 4} \approx 1.83 \,.
\end{equation}
\end{itemize}
For the cases we have considered above, we have the following
\begin{equation}\label{eq:frequency_ranges}
1.46 \lesssim b_{\ell = 3} \lesssim 1.64 \,, \hspace{2cm} 1.83 \lesssim b_{\ell = 4} \lesssim 2.18 \,.
\end{equation}

If values are known for the magnitude and frequency of the different GW $\ell$-modes, then precise estimates can be made for the GW damping and heating effects. However, here we adopt a more heuristic approach, and use the frequencies given above to investigate the significance of the effects. Since the frequency varies with $\ell$, the GW damping effect per $\ell$-mode is, in principle, directly observable; whereas an observation of a temperature increase (of matter close to a GW source) would not provide any information about the contribution to the heating by the various $\ell$-modes. Thus, only the GW damping effect is analyzed further. 

It is clear from Eq.~\eqref{e-HD} that the GW damping effect for the $\ell=3,4$ modes is stronger than for the $\ell=2$ mode whenever $D_{\ell=3,4}>D_{\ell=2}$, and that case applies to the astrophysical examples considered below. However, it should be noted that there are regions in the parameter space for which $D_{\ell=3,4}<D_{\ell=2}$: e.g., setting $\alpha=2$, $r_i=15$km, $f=5$kHz, $b_3=1.60$, and $b_4=2.17$ leads to $D_2=3.61$, $D_3=2.95$, and $D_4=2.80$.

We now compare the damping factors in the $\ell = 3$ and $\ell = 4$ cases to the $\ell = 2$ case for two astrophysical scenarios. In both cases, we use a fixed value of $\alpha=2$, i.e., $r_o=2r_i$ because the GW damping effect is insensitive to the value of $\alpha$~\citep{Bishop:2022kzq}. In the figures below, damping factors are plotted against the frequency of the $\ell = 2$ mode; thus for $\ell=3,4$ the actual frequency is obtained after multiplication by $b_\ell$. 

The first case we consider is that of the post-merger signal from a BNS merger, generated by the quasinormal modes of a hypermassive neutron star (HMNS) remnant. The astrophysical parameters for this scenario were discussed in detail in~\cite{Bishop:2024oht}, which derived the damping effect of the $\ell=2$ mode, and are: $r_i \approx 12$km, and $1$kHz $\lesssim f_{\ell=2} \lesssim 2$kHz. Since the damping ratios are sensitive to the choice of the inner radius, we shall consider $r_i$ values of $12$km and $15$km. Eq.~\eqref{eq:HNS_frequencies} is used for the frequency ratios. The value used for the shear viscosity is $\eta = 10^{24}$kg/m/s, which is at the low end of its expected value~\cite{Bishop:2024oht}; thus the results presented below may underestimate the effect.

\begin{figure}[h]
    \centering
    \includegraphics[width=0.48\textwidth]{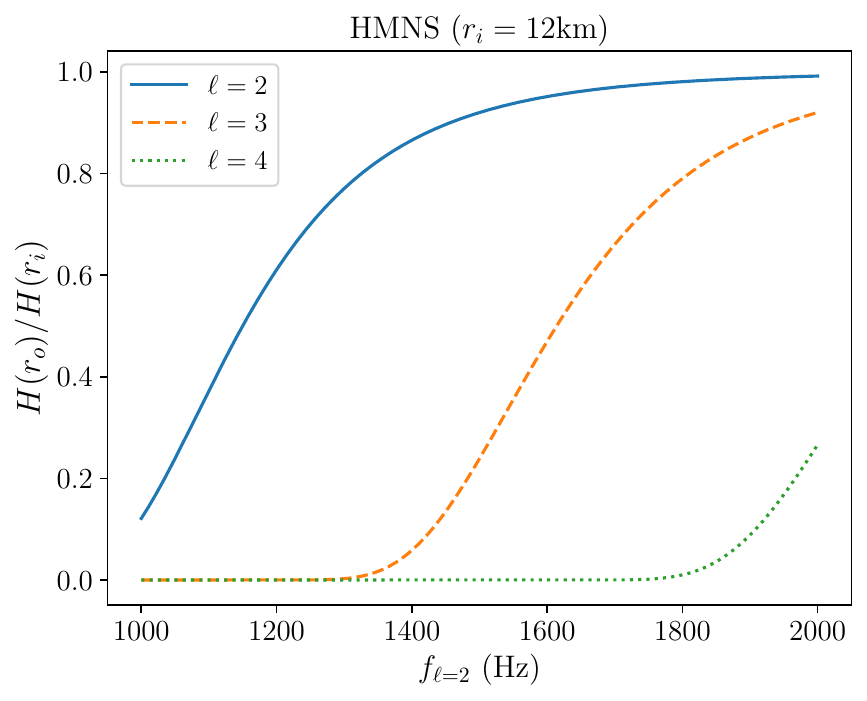}
    \hfill
    \includegraphics[width=0.48\textwidth]{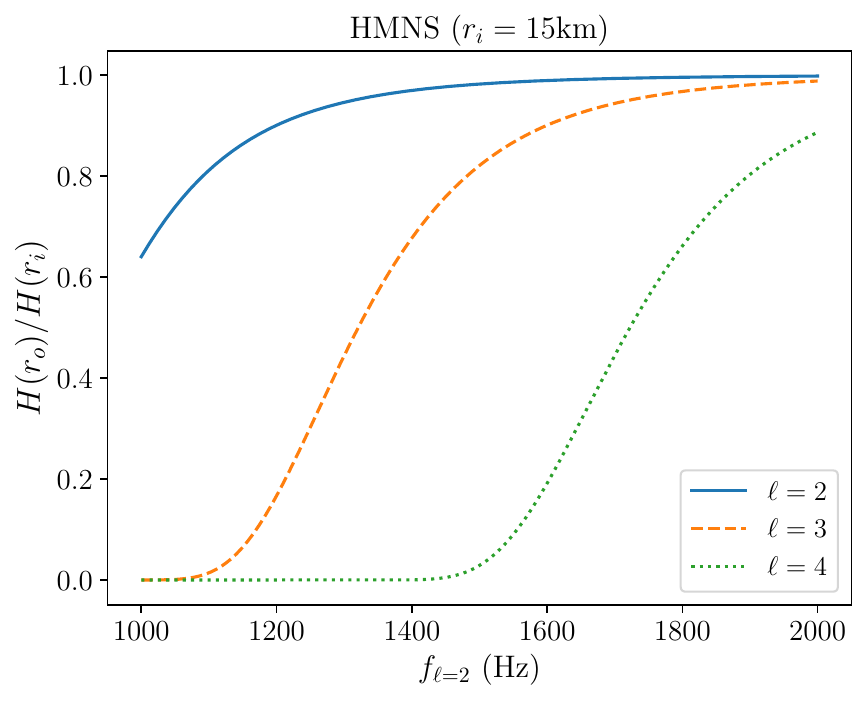}
    \caption{Plots of the damping factors when the $\ell$-modes are either $2$, $3$, or $4$, for the case of an HMNS. The left panel was obtained by using an inner radius value of $r_i = 12$km, whereas the right panel was obtained by setting $r_i = 15$km. For both plots, we have used $\eta = 10^{24}$kg/m/s. We have also used Eq.~\eqref{eq:HNS_frequencies} for the frequency ratios associated with the $\ell = 3$ and $4$ cases.}
    \label{fig:fig_HMNS}
\end{figure}

In Fig.~\ref{fig:fig_HMNS}, we plot the damping factors for $\ell$-modes of $2$, $3$, and $4$, for the case of an HMNS. The plots show that higher $\ell$-modes are always more strongly damped than in the $\ell=2$ case. The magnitude of the damping depends on the parameters $f$, $r_i$, and $\ell$, and varies between minimal damping to almost complete damping. It should be noted that the damping of the $\ell=3,4$ modes is almost complete at frequency values towards the bottom of the range.

\begin{figure}[h]
    \centering
    \includegraphics[width=0.48\textwidth]{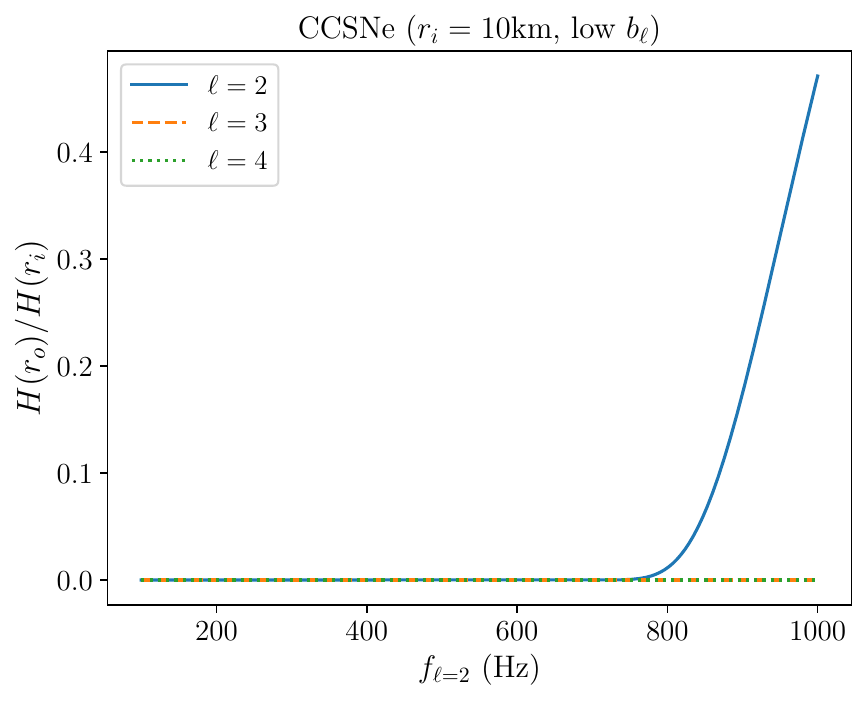}
    \hfill
    \includegraphics[width=0.48\textwidth]{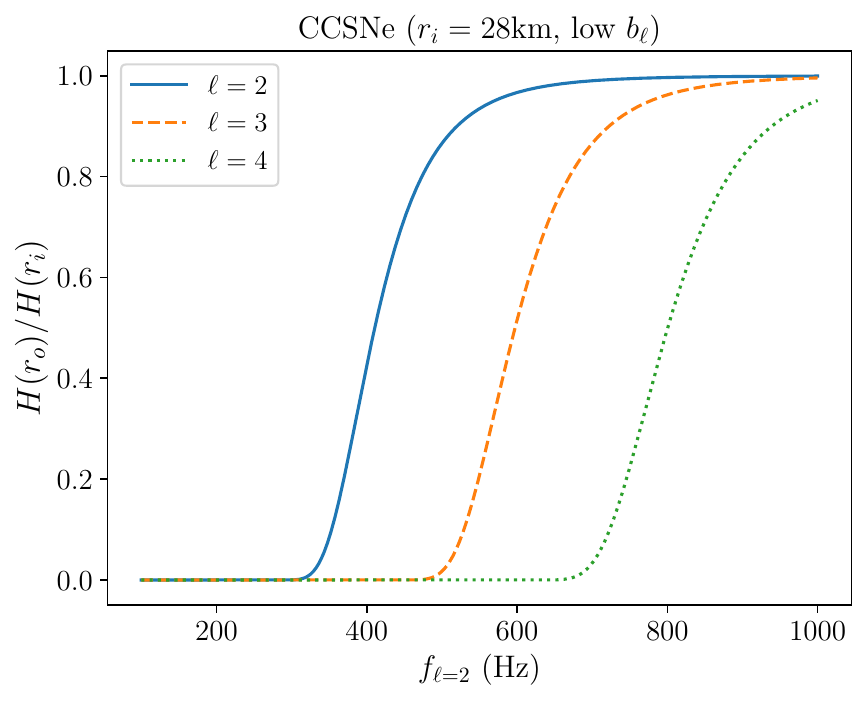}
    \\
        \includegraphics[width=0.48\textwidth]{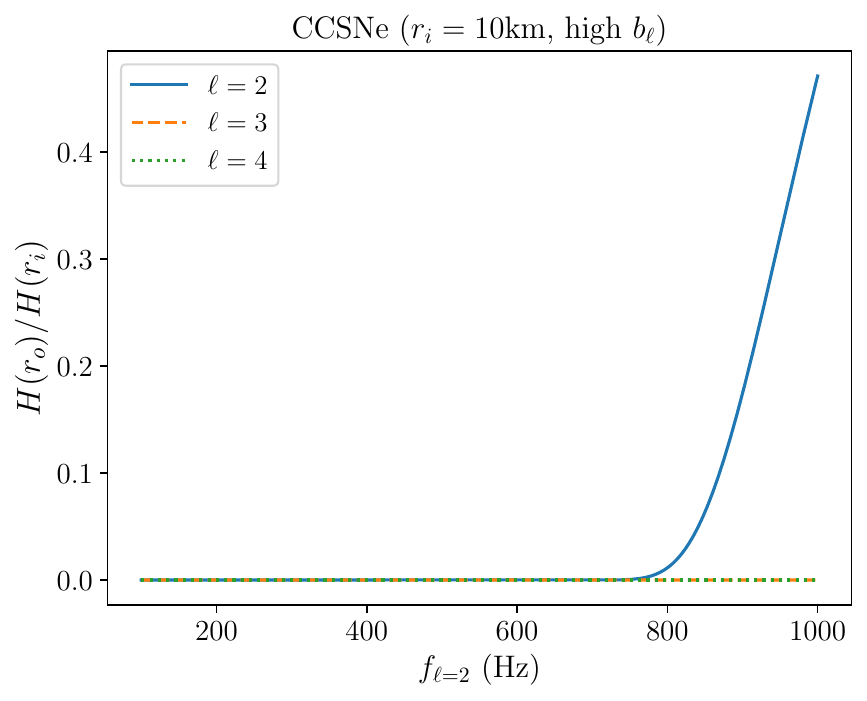}
    \hfill
    \includegraphics[width=0.48\textwidth]{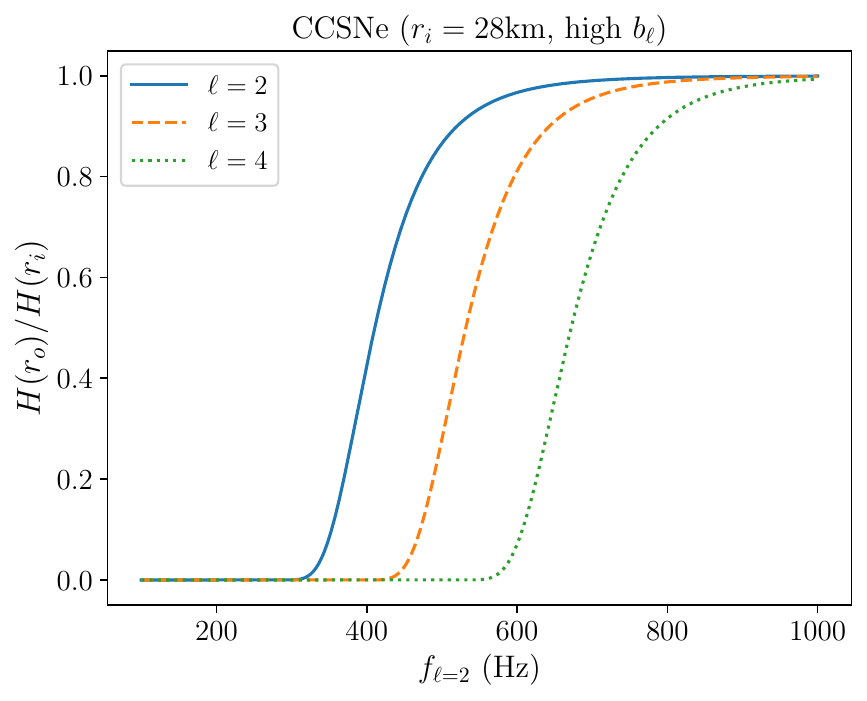}
    \caption{Plots of the damping factors of various $\ell$-modes against the frequency of the $\ell=2$ mode, for the case of CCSNe. All the plots use $\eta = 10^{23}$kg/m/s. The top left panel has $r_i=10$km and low  $b_\ell$ ($b_{\ell=3}=1.46,b_{\ell=4}=1.83$), the top right panel has  $r_i=28$km and low $b_\ell$, the bottom left panel has $r_i=10$km and high $b_\ell$ ($b_{\ell=3}=1.64,b_{\ell=4}=2.18$), and the bottom right panel has $r_i=28$km and high $b_\ell$.}
    \label{fig:fig_CCSNe}
\end{figure}

The second astrophysical scenario we consider is that of a CCSNe. Such a scenario was considered in~\cite{Bishop:2022kzq} for the $\ell = 2$ case. The astrophysical parameters used were $100$Hz$\lesssim f_{\ell=2} \lesssim 1$kHz, and $10$km$\lesssim r_i \lesssim 28$km. The smallest value considered for the viscosity was $\eta = 10^{23}$kg/m/s, and that value will be used in the calculations below; thus, again, the results obtained may be an underestimate of the magnitude of the GW damping effect. There is little information in the literature about the frequencies of higher order $\ell$-modes, so we use the range of values given in Eq.~\eqref{eq:frequency_ranges}: we will call $b_{\ell=3}=1.46$, $b_{\ell=4}=1.83$ \textit{low} $b_\ell$, and $b_{\ell=3}=1.64$, $b_{\ell=4}=2.18$ \textit{high} $b_\ell$. We consider the two cases $r_i = 28$km and $r_i=10$km. For these parameters, we plot the damping factors in Fig.~\ref{fig:fig_CCSNe}. As was found for the HMNS case, higher $\ell$-modes are always more strongly damped than for the $\ell=2$ mode. The magnitude of the damping varies between minimal damping to almost complete damping, and the damping of the $\ell=3,4$ modes is almost complete at frequency values or inner radii ($r_i$) towards the bottom of the expected range.

\section{Summary and Conclusion}\label{sec:conclusions}
Previous work on GWs on a Minkowski background within the Bondi--Sachs framework was mainly limited to quadrupolar perturbations, i.e., to the case that angular dependence of the GWs is described by an $\ell=2$ spherical harmonic. This work has extended these results to perturbations with an arbitrary value of $\ell\ge 2$ for the spherical harmonic, and expressions have been obtained for the metric components, as well as for the velocity and shear of a matter field around the GW source. These expressions have then been used to evaluate the GW damping and heating effects when the matter is viscous. It was found that, for constant frequency, the effects are enhanced as $\ell$ increases. Allowing the frequency to increase with $\ell$, as is expected in astrophysical scenarios, the effect is still enhanced for long wavelengths, i.e., when $\psi_{\ell=2}=\lambda_{\ell=2}/(2\pi r_i)$ is somewhat greater than unity.

When $\ell=2$, the GW damping and heating effects are expected to be significant for the astrophysical scenarios of BNS and CCSNe. We investigated these scenarios when $\ell=3$ and $4$, and found that the effects are always enhanced, perhaps substantially so depending on the actual values of some uncertain astrophysical parameters. In particular, it was found that the damping of the $\ell=3,4$ modes is almost complete when the frequency is towards the lower end of the expected range. If a nearby BNS or CCSNe occurs, then this work indicates that higher order $\ell$ modes would not be present in the GW signal, except possibly at frequencies towards the top of the expected range; further, in the BNS case, such a signal would be outside the frequency sensitivity of current detectors and so would not be observed.

It has been shown when $\ell=2$ that use of the more physically appropriate Schwarzschild background can produce GW damping and heating effects larger by a factor ${\mathcal O}(10)$ than on a Minkowski background. To what extent this applies to higher $\ell$ modes is unknown; this would need to be investigated numerically, and is deferred to further work.

	\begin{acknowledgments}
		This work was supported by the National Research Foundation, South Africa, under grant number CPRR240314209194.
	\end{acknowledgments}
	\section*{Conflict of interest}
	%%%%%%%%%%%%%%%%%%%%%%%%%%%%%%%%%%%%%%%%%%%%
	The authors declare that they have no conflict of interest.

\appendix
\section{Linearized field equations}\label{sec:linearised_field_equations}
The linearized field equations within the Bondi--Sachs framework are given as follows:
\begin{align}
\frac{4 \partial_r \beta}{r} = 0 \,, \label{eq:R11} \\
\frac{r^{2} \partial_r^2 U}{2} + 2 r \partial_r U + \frac{\partial_r \bar{\eth} J}{2} - \partial_r \eth \beta + \frac{2 \eth \beta}{r} = 0 \,, \label{eq:qA_RA1} \\
- r^{2} \partial_r^2 J + r^{2} \partial_r \eth U + 2 r^{2} \partial_u \partial_r J + 2 r \eth U - 2 r \partial_r J + 2 r \partial_u J - 2 \eth^2 \beta = 0 \,, \label{eq:RJ} \\
\frac{\partial_r \bar{\eth}U}{2} + \frac{\partial_r \eth\bar{U}}{2} + \frac{2 \bar{\eth}U}{r} + \frac{2 \eth\bar{U}}{r} - \frac{2 \bar{\eth} \eth \beta}{r^{2}} + \frac{\bar{\eth}^2 J}{2 r^{2}} + \frac{4 \beta}{r^{2}} + \frac{\eth^2 \bar{J}}{2 r^{2}} - \frac{2 \partial_r W}{r^{2}} = 0 \,, \label{eq:hAB_RAB} \\
- \frac{\partial_u \bar{\eth}U}{2} + \partial_r^2 \beta - \frac{\partial_u \eth\bar{U}}{2} - 2 \partial_u \partial_r \beta + \frac{\partial_r^2 W}{2 r} + \frac{2 \partial_r \beta}{r} - \frac{2 \partial_u \beta}{r} + \frac{\bar{\eth} \eth \beta}{r^{2}} + \frac{\partial_u W}{r^{2}} + \frac{\bar{\eth} \eth W}{2 r^{3}} = 0 \,, \label{eq:R00} \\
- \frac{\partial_r \bar{\eth}U}{4} + \partial_r^2 \beta - \frac{\partial_r \eth\bar{U}}{4} - 2 \partial_u \partial_r \beta - \frac{\bar{\eth}U}{2 r} - \frac{\eth\bar{U}}{2 r} + \frac{\partial_r^2 W}{2 r} + \frac{2 \partial_r \beta}{r} + \frac{\bar{\eth} \eth \beta}{r^{2}} = 0 \,, \label{eq:R01} \\
\frac{r^{2} \partial_r^2 U}{2} - \frac{r^{2} \partial_u \partial_r U}{2} + 2 r \partial_r U + U + \frac{\eth \bar{\eth} U}{4} - \frac{\eth^2 \bar{U}}{4} + \frac{\partial_u \bar{\eth} J}{2} - \partial_u \eth \beta + \frac{\partial_r \eth W}{2 r} - \frac{\eth W}{2 r^{2}} = 0 \,, \label{eq:qA_RA0}
\end{align}
which are, respectively, the linearized $R_{11}$, $q^A R_{A 1}$, $q^A q^B R_{AB}$, $h^{A B} R_{A B}$, $R_{0 0}$, $R_{0 1}$, and $q^A R_{A 0}$ field equations. We note that such equations have been studied previously in~\cite{Bishop:2004ug}, and we have provided them here for convenience.
\section{Derivation of Eq.~\eqref{eq:sigma_squared}}\label{sec:derivation_of_sigma_squared}
As noted in the main text, $B_{(0a)} = 0$ and thus, for the GW solutions considered here, we have $\sigma_{0a} = 0$. We now have the following linearized expression
\begin{align}
\sigma_{ab} \sigma^{ab} = \sigma_{11}^2 + 2 \eta^{AB} \sigma_{1A} \sigma_{1B} + \eta^{AC} \eta^{BD} \sigma_{AB} \sigma_{CD} \,.
\end{align}
Using $\eta^{AB} = q^{AB} / r^2$, we have
\begin{align}\label{eq:sigma_squared_intermediate_1}
\sigma_{ab} \sigma^{ab} = \sigma_{11}^2 + \tfrac{2}{r^2} q^{AB} \sigma_{1A} \sigma_{1B} + \tfrac{1}{r^4} q^{AC} q^{BD} \sigma_{AB} \sigma_{CD} \,,
\end{align}
and recall that $q_{AB}$ is the metric for the unit $2$-sphere. The complex dyad is related to this metric through $q_{AB} = q_{(A} \bar{q}_{B)}$ which gives us the following after substituting into Eq.~\eqref{eq:sigma_squared_intermediate_1}
\begin{align}\label{eq:sigma_squared_intermediate_2}
\sigma_{ab} \sigma^{ab} = \sigma_{11}^2 + \tfrac{2}{r^2} q^A \bar{q}^B \sigma_{1A} \sigma_{1B} + \tfrac{1}{2r^4} q^A \bar{q}^C q^B \bar{q}^D \sigma_{AB} \sigma_{CD} + \tfrac{1}{2r^4} q^A \bar{q}^C \bar{q}^B q^D \sigma_{AB} \sigma_{CD} \,.
\end{align}
By making use of $q^A \bar{q}^B \sigma_{AB} / r^2 = -\sigma_{11}$, we see that the last term is nothing more than $\sigma_{11}^2 / 2$. It follows that Eq.~\eqref{eq:sigma_squared_intermediate_2} reduces to Eq.~\eqref{eq:sigma_squared}.
\section{Metric and matter variables for $\ell = 2$, $3$, and $4$}\label{sec:specific_ell_metric}
In this section, we provide the radial parts of the metric components, velocity field components, and shear tensor components when $\ell = 2$, $3$, or $4$. We also give expressions for $\mathcal{N}_{\ell, \ell}$ and $\langle\dot{E}_{\text{GW}}\rangle$ for these values of $\ell$ by making use of Eqs.~\eqref{eq:News} and~\eqref{eq:E_GW}, respectively. 

The ansatz for the metric variables are given in Eq.~\eqref{eq:metric_ansatz} while the ansatz used for the shear tensor is given in Eq.~\eqref{eq:shear:ansatz}. For the four-velocity, we have used the following
\begin{align}\label{eq:velocity_fields_ansatz}
V_0 &= -1 + \Re\left(V_0^{[\ell, \ell]}(r) {\text e}^{i u \nu}\right) \,_0 Z_{\ell, \ell} \,, \hspace{2cm} V_1 = -1 + \Re\left(V_1^{[\ell, \ell]}(r) {\text e}^{i u \nu}\right) \,_0 Z_{\ell, \ell} \,, \nonumber \\
V_{\text{ang}} &= \Re\left(V_{\text{ang}}^{[\ell, \ell]}(r) {\text e}^{i u \nu}\right) \,_1 Z_{\ell, \ell} \,,
\end{align}
Specific $\ell$ expressions for $U^{[\ell, \ell]}$, $J^{[\ell, \ell]}$, and $W^{[\ell, \ell]}$ are given here by, respectively, using Eqs.~\eqref{eq:general_U_solution}, \eqref{eq:general_J_solution}, and ~\eqref{eq:general_W_solution}. In addition, the expressions given in this section for the shear tensor components are obtained by making use of Eqs.~\eqref{eq:sigma_11_ell}--\eqref{eq:sigma_U_ell}. Expressions for the radial parts of the velocity field can be obtained by substituting the expressions for the metric components, as well as the ansatz for the velocity field components given in Eq.~\eqref{eq:velocity_fields_ansatz}, into Eqs.~\eqref{eq:V0_eths} and~\eqref{eq:V1_ang_eths}.
\subsection{$\ell = 2$}
The metric components for the $\ell = 2$ case have been discussed previously in~\cite{Bishop:2004ug, Bishop:2019ckc, Bishop:2022kzq}. Using the arbitrary $\ell$ expressions given in the main text, we find
\begin{align}
J^{[2, 2]} &= \frac{2 \sqrt{6} i C_{30} \nu}{3} + \frac{2 \sqrt{6} C_{30}}{r} + \frac{2 \sqrt{6} i C_{40} \nu^{3}}{3} + \frac{2 \sqrt{6} C_{40}}{r^{3}} + \frac{4 \sqrt{6} b_{0}}{3} \,, \\
U^{[2, 2]} &= \frac{\sqrt{6} C_{30} \nu^{2}}{3} + \frac{2 \sqrt{6} C_{30}}{r^{2}} + \frac{\sqrt{6} C_{40} \nu^{4}}{3} - \frac{4 \sqrt{6} i C_{40} \nu}{r^{3}} - \frac{3 \sqrt{6} C_{40}}{r^{4}} - \frac{2 \sqrt{6} i \nu b_{0}}{3} + \frac{2 \sqrt{6} b_{0}}{r} \,, \\
W^{[2, 2]} &= - 2 C_{30} \nu^{2} r^{2} + 4 i C_{30} \nu r - 2 C_{40} \nu^{4} r^{2} + 4 i C_{40} \nu^{3} r + 12 C_{40} \nu^{2} - \frac{12 i C_{40} \nu}{r} - \frac{6 C_{40}}{r^{2}} + 4 i \nu b_{0} r^{2} - 2 b_{0} r \,.
\end{align}
The expressions given here for $J^{[2, 2]}$ and $W^{[2, 2]}$ match those given in~\cite{Bishop:2022kzq} (see Eq.~(3) therein), and we correct a misprint in the $-4 i \nu \sqrt{6} C_{40}$ term in the expression for $U^{[2, 2]}$, which should have a $1/r^3$ dependence.

For the radial parts of the velocity field components, we find
\begin{align}
V_0^{[2, 2]} &= C_{30} \nu^{2} r - 2 i C_{30} \nu + C_{40} \nu^{4} r - 2 i C_{40} \nu^{3} - \frac{6 C_{40} \nu^{2}}{r} + \frac{6 i C_{40} \nu}{r^{2}} + \frac{3 C_{40}}{r^{3}} - 2 i \nu b_{0} r \,, \\
V_1^{[2, 2]} &= - i C_{30} \nu - i C_{40} \nu^{3} - \frac{6 i C_{40} \nu}{r^{2}} - \frac{12 C_{40}}{r^{3}} + \frac{9 i C_{40}}{\nu r^{4}} - 2 b_{0} \,, \\
V_{\text{ang}}^{[2, 2]} &= - \sqrt{6} i C_{30} \nu r - 2 \sqrt{6} C_{30} - \sqrt{6} i C_{40} \nu^{3} r - 2 \sqrt{6} C_{40} \nu^{2} + \frac{6 \sqrt{6} i C_{40} \nu}{r} + \frac{6 \sqrt{6} C_{40}}{r^{2}} - \frac{3 \sqrt{6} i C_{40}}{\nu r^{3}} - 2 \sqrt{6} b_{0} r \,,
\end{align}
which are the same as the expressions given in Eq.~(6) of~\cite{Bishop:2022kzq}.

For the radial parts of the shear tensor components, we have
\begin{align}
\sigma^{[2, 2]}_{11} &= \frac{12 i C_{40} \nu}{r^{3}} + \frac{36 C_{40}}{r^{4}} - \frac{36 i C_{40}}{\nu r^{5}} \,, \\
\sigma^{[2, 2]}_{J} / r^2 &= - \frac{2 \sqrt{6} i C_{40} \nu^{3}}{r} - \frac{4 \sqrt{6} C_{40} \nu^{2}}{r^{2}} + \frac{6 \sqrt{6} i C_{40} \nu}{r^{3}} + \frac{6 \sqrt{6} C_{40}}{r^{4}} - \frac{6 \sqrt{6} i C_{40}}{\nu r^{5}} \,, \\
\sigma^{[2, 2]}_{U} / r &= \frac{2 \sqrt{6} C_{40} \nu^{2}}{r^{2}} - \frac{6 \sqrt{6} i C_{40} \nu}{r^{3}} - \frac{12 \sqrt{6} C_{40}}{r^{4}} + \frac{12 \sqrt{6} i C_{40}}{\nu r^{5}} \,.
\end{align}
The expression for $\sigma^{[2, 2]}_{11}$ given here matches the expression given in Eq.~(9) of~\cite{Bishop:2022kzq}. We note that $\sigma^{[2, 2]}_U$ given in~\cite{Bishop:2022kzq} should have a factor of $\sqrt{6}$, and the expression for $\sigma^{[2, 2]}_J$ should have a factor of $2 i \sqrt{6}$. The expressions given here reflect these corrections. We emphasize that the expressions for the rescaled GW magnitude given in Eq.~\eqref{eq:H_ell_2} matches the one given in~\cite{Bishop:2022kzq}.

For the news and the rate of energy that is output as GWs, we have
\begin{equation}\label{eq:news_E_GW_ell_2}
N_{2, 2} = \sqrt{6} \nu^3 \Re\left(-iC_{40} {\text e}^{i \nu u}\right) \,_2 Z_{2, 2} \,, \hspace{3cm} \langle{\dot{E}_{\text{GW}}}\rangle_{\ell = 2} = \frac{3 |C_{40}|^{2} \nu^{6}}{4 \pi} \,,
\end{equation}
which have been reported previously~\cite{Bishop:2022kzq}.
\subsection{$\ell = 3$}\label{sec:ell_3_expressions}
The metric variables for the $\ell = 3$ case have been studied previously in~\cite{Reisswig:2006nt} (see Eq.~(14) therein). Using Eqs.~\eqref{eq:general_J_solution}, \eqref{eq:general_U_solution}, and~\eqref{eq:general_W_solution} of the present work, we find
\begin{align}
J^{[3, 3]} &= \frac{2 \sqrt{30} i C_{30} \nu}{15} + \frac{4 \sqrt{30} C_{30}}{5 r} + \frac{2 \sqrt{3} i C_{40} \nu^{3}}{15} + \frac{4 \sqrt{3} C_{40}}{r^{3}} - \frac{6 \sqrt{3} i C_{40}}{\nu r^{4}} + \frac{2 \sqrt{30} b_{0}}{3} \,, \\
U^{[3, 3]} &= \frac{2 \sqrt{3} C_{30} \nu^{2}}{15} + \frac{4 \sqrt{3} C_{30}}{r^{2}} + \frac{\sqrt{30} C_{40} \nu^{4}}{75} - \frac{8 \sqrt{30} i C_{40} \nu}{5 r^{3}} - \frac{3 \sqrt{30} C_{40}}{r^{4}} + \frac{12 \sqrt{30} i C_{40}}{5 \nu r^{5}} - \frac{2 \sqrt{3} i \nu b_{0}}{3} + \frac{4 \sqrt{3} b_{0}}{r} \,, \\
W^{[3, 3]} &= - \frac{4 C_{30} \nu^{2} r^{2}}{5} + 4 i C_{30} \nu r - \frac{2 \sqrt{10} C_{40} \nu^{4} r^{2}}{25} + \frac{2 \sqrt{10} i C_{40} \nu^{3} r}{5} + \frac{12 \sqrt{10} C_{40} \nu^{2}}{5} - \frac{24 \sqrt{10} i C_{40} \nu}{5 r} - \frac{6 \sqrt{10} C_{40}}{r^{2}} \nonumber\\
&+ \frac{18 \sqrt{10} i C_{40}}{5 \nu r^{3}} + 4 i \nu b_{0} r^{2} - 2 b_{0} r \,.
\end{align}
The expressions given in~\cite{Reisswig:2006nt} use a different notation, which we now compare to the notation used here. The constants $\beta_0$, $C_1$ and $C_2$ used in~\cite{Reisswig:2006nt} are related to the constants used in the present work through $\beta_0 = b_0$, $C_{30} = C_1 / 4$, and $C_{40} = -i \nu \sqrt{10} C_2 / 12$. In addition, the $U_3$ and $J_3$ used in~\cite{Reisswig:2006nt} are such that $U^{[3, 3]} = 2 \sqrt{3} U_3$ and $J^{[3, 3]} = 2 \sqrt{30} J_3$. With these notational differences, we find that the $U$ and $J$ expressions match. Regarding the radial part of $W$, which is expressed as $W_{c} = W / r^2$ in~\cite{Reisswig:2006nt}, we note that the expressions match if we note the following corrections. On the right-hand side of the third equality in Eq.~(14) of~\cite{Reisswig:2006nt}, the $3r$ denominator should be $r$. In addition, a factor of $i$ should be absent in the third to last term, and present in the second to last term.

Below, we now provide the velocity field and shear tensor components for the case where $\ell = 3$:
\begin{align}
V_0^{[3, 3]} &= \frac{2 C_{30} \nu^{2} r}{5} - 2 i C_{30} \nu + \frac{\sqrt{10} C_{40} \nu^{4} r}{25} - \frac{\sqrt{10} i C_{40} \nu^{3}}{5} - \frac{6 \sqrt{10} C_{40} \nu^{2}}{5 r} + \frac{12 \sqrt{10} i C_{40} \nu}{5 r^{2}} + \frac{3 \sqrt{10} C_{40}}{r^{3}} \nonumber\\
&- \frac{9 \sqrt{10} i C_{40}}{5 \nu r^{4}} - 2 i \nu b_{0} r \,, \\
V_1^{[3, 3]} &= - \frac{2 i C_{30} \nu}{5} - \frac{\sqrt{10} i C_{40} \nu^{3}}{25} - \frac{6 \sqrt{10} i C_{40} \nu}{5 r^{2}} - \frac{24 \sqrt{10} C_{40}}{5 r^{3}} + \frac{9 \sqrt{10} i C_{40}}{\nu r^{4}} + \frac{36 \sqrt{10} C_{40}}{5 \nu^{2} r^{5}} - 2 b_{0} \,,\\
V_{\text{ang}}^{[3, 3]} &= - \frac{4 \sqrt{3} i C_{30} \nu r}{5} - 4 \sqrt{3} C_{30} - \frac{2 \sqrt{30} i C_{40} \nu^{3} r}{25} - \frac{2 \sqrt{30} C_{40} \nu^{2}}{5} + \frac{12 \sqrt{30} i C_{40} \nu}{5 r} + \frac{24 \sqrt{30} C_{40}}{5 r^{2}} - \frac{6 \sqrt{30} i C_{40}}{\nu r^{3}} \nonumber\\
&- \frac{18 \sqrt{30} C_{40}}{5 \nu^{2} r^{4}} - 4 \sqrt{3} b_{0} r \,, \\
\sigma^{[3, 3]}_{11} &= \frac{12 \sqrt{10} i C_{40} \nu}{5 r^{3}} + \frac{72 \sqrt{10} C_{40}}{5 r^{4}} - \frac{36 \sqrt{10} i C_{40}}{\nu r^{5}} - \frac{36 \sqrt{10} C_{40}}{\nu^{2} r^{6}} \,, \\
\sigma^{[3, 3]}_{J} / r^2 &= - \frac{4 \sqrt{3} i C_{40} \nu^{3}}{5 r} - \frac{4 \sqrt{3} C_{40} \nu^{2}}{r^{2}} + \frac{12 \sqrt{3} i C_{40} \nu}{r^{3}} + \frac{24 \sqrt{3} C_{40}}{r^{4}} - \frac{36 \sqrt{3} i C_{40}}{\nu r^{5}} - \frac{36 \sqrt{3} C_{40}}{\nu^{2} r^{6}} \,, \\
\sigma^{[3, 3]}_{U} / r &= \frac{2 \sqrt{30} C_{40} \nu^{2}}{5 r^{2}} - \frac{12 \sqrt{30} i C_{40} \nu}{5 r^{3}} - \frac{42 \sqrt{30} C_{40}}{5 r^{4}} + \frac{18 \sqrt{30} i C_{40}}{\nu r^{5}} + \frac{18 \sqrt{30} C_{40}}{\nu^{2} r^{6}} \,.
\end{align}

For $\mathcal{N}_{\ell, \ell}$ and $\langle\dot{E}_{\text{GW}}\rangle$, we find
\begin{equation}
N_{3, 3} = \frac{6 \nu^3 \Re\left(-iC_{40} {\text e}^{i \nu u}\right)}{5 \sqrt{3}} \,_2 Z_{3, 3} \,, \hspace{3cm} \langle{\dot{E}_{\text{GW}}}\rangle_{\ell = 3} = \frac{3 |C_{40}|^{2} \nu^{6}}{50 \pi} \,.
\end{equation}
As noted in the main text, the news given here for the $\ell=3$ case agrees with Eq.(16) of~\cite{Reisswig:2006nt}.
\subsection{$\ell = 4$}
The metric variables, velocity field components, and shear tensor components for the $\ell = 4$ case are
\begin{align}
J^{[4, 4]} &= \frac{2 \sqrt{10} i C_{30} \nu}{15} + \frac{4 \sqrt{10} C_{30}}{3 r} + \frac{2 \sqrt{5} i C_{40} \nu^{3}}{75} + \frac{4 \sqrt{5} C_{40}}{r^{3}} - \frac{14 \sqrt{5} i C_{40}}{\nu r^{4}} - \frac{84 \sqrt{5} C_{40}}{5 \nu^{2} r^{5}} + \frac{6 \sqrt{10} b_{0}}{5} \,, \\
U^{[4, 4]} &= \frac{2 \sqrt{5} C_{30} \nu^{2}}{45} + \frac{4 \sqrt{5} C_{30}}{r^{2}} + \frac{\sqrt{10} C_{40} \nu^{4}}{225} - \frac{8 \sqrt{10} i C_{40} \nu}{3 r^{3}} - \frac{9 \sqrt{10} C_{40}}{r^{4}} + \frac{84 \sqrt{10} i C_{40}}{5 \nu r^{5}} + \frac{14 \sqrt{10} C_{40}}{\nu^{2} r^{6}} \nonumber\\
&- \frac{2 \sqrt{5} i \nu b_{0}}{5} + \frac{4 \sqrt{5} b_{0}}{r} \,, \\
W^{[4, 4]} &= - \frac{4 C_{30} \nu^{2} r^{2}}{9} + 4 i C_{30} \nu r - \frac{2 \sqrt{2} C_{40} \nu^{4} r^{2}}{45} + \frac{2 \sqrt{2} i C_{40} \nu^{3} r}{5} + 4 \sqrt{2} C_{40} \nu^{2} - \frac{40 \sqrt{2} i C_{40} \nu}{3 r} - \frac{30 \sqrt{2} C_{40}}{r^{2}} \nonumber\\
&+ \frac{42 \sqrt{2} i C_{40}}{\nu r^{3}} + \frac{28 \sqrt{2} C_{40}}{\nu^{2} r^{4}} + 4 i \nu b_{0} r^{2} - 2 b_{0} r \,, \\
V_0^{[4, 4]} &= \frac{2 C_{30} \nu^{2} r}{9} - 2 i C_{30} \nu + \frac{\sqrt{2} C_{40} \nu^{4} r}{45} - \frac{\sqrt{2} i C_{40} \nu^{3}}{5} - \frac{2 \sqrt{2} C_{40} \nu^{2}}{r} + \frac{20 \sqrt{2} i C_{40} \nu}{3 r^{2}} + \frac{15 \sqrt{2} C_{40}}{r^{3}} - \frac{21 \sqrt{2} i C_{40}}{\nu r^{4}} \nonumber\\
&- \frac{14 \sqrt{2} C_{40}}{\nu^{2} r^{5}} - 2 i \nu b_{0} r \,, \\
V_1^{[4, 4]} &= - \frac{2 i C_{30} \nu}{9} - \frac{\sqrt{2} i C_{40} \nu^{3}}{45} - \frac{2 \sqrt{2} i C_{40} \nu}{r^{2}} - \frac{40 \sqrt{2} C_{40}}{3 r^{3}} + \frac{45 \sqrt{2} i C_{40}}{\nu r^{4}} + \frac{84 \sqrt{2} C_{40}}{\nu^{2} r^{5}} - \frac{70 \sqrt{2} i C_{40}}{\nu^{3} r^{6}} - 2 b_{0} \,, \\
V_{\text{ang}}^{[4, 4]} &= - \frac{4 \sqrt{5} i C_{30} \nu r}{9} - 4 \sqrt{5} C_{30} - \frac{2 \sqrt{10} i C_{40} \nu^{3} r}{45} - \frac{2 \sqrt{10} C_{40} \nu^{2}}{5} + \frac{4 \sqrt{10} i C_{40} \nu}{r} + \frac{40 \sqrt{10} C_{40}}{3 r^{2}} - \frac{30 \sqrt{10} i C_{40}}{\nu r^{3}} \nonumber\\
&- \frac{42 \sqrt{10} C_{40}}{\nu^{2} r^{4}} + \frac{28 \sqrt{10} i C_{40}}{\nu^{3} r^{5}} - 4 \sqrt{5} b_{0} r \,, \\
\sigma^{[4, 4]}_{11} &= \frac{4 \sqrt{2} i C_{40} \nu}{r^{3}} + \frac{40 \sqrt{2} C_{40}}{r^{4}} - \frac{180 \sqrt{2} i C_{40}}{\nu r^{5}} - \frac{420 \sqrt{2} C_{40}}{\nu^{2} r^{6}} + \frac{420 \sqrt{2} i C_{40}}{\nu^{3} r^{7}} \,, \\
\sigma^{[4, 4]}_{J} / r^2 &= - \frac{4 \sqrt{5} i C_{40} \nu^{3}}{15 r} - \frac{12 \sqrt{5} C_{40} \nu^{2}}{5 r^{2}} + \frac{12 \sqrt{5} i C_{40} \nu}{r^{3}} + \frac{40 \sqrt{5} C_{40}}{r^{4}} - \frac{96 \sqrt{5} i C_{40}}{\nu r^{5}} - \frac{168 \sqrt{5} C_{40}}{\nu^{2} r^{6}} + \frac{168 \sqrt{5} i C_{40}}{\nu^{3} r^{7}} \,, \\
\sigma^{[4, 4]}_{U} / r &= \frac{2 \sqrt{10} C_{40} \nu^{2}}{5 r^{2}} - \frac{4 \sqrt{10} i C_{40} \nu}{r^{3}} - \frac{22 \sqrt{10} C_{40}}{r^{4}} + \frac{78 \sqrt{10} i C_{40}}{\nu r^{5}} + \frac{168 \sqrt{10} C_{40}}{\nu^{2} r^{6}} - \frac{168 \sqrt{10} i C_{40}}{\nu^{3} r^{7}} \,.
\end{align}
Using Eqs.~\eqref{eq:News} and~\eqref{eq:E_GW}, we find the following
\begin{equation}
N_{4, 4} = \frac{2 \nu^3 \Re\left(-iC_{40} {\text e}^{i \nu u}\right)}{3 \sqrt{5}} \,_2 Z_{4, 4} \,, \hspace{3cm} \langle{\dot{E}_{\text{GW}}}\rangle_{\ell = 4} = \frac{|C_{40}|^{2} \nu^{6}}{90 \pi} \,.
\end{equation}
\section{Applicability of using linearized expressions and a test shell approximation}\label{sec:validity_discussion}
The results given in the present work were obtained by using the linearized field equations. In addition, back-reaction effects of the matter shell onto the metric were taken to be negligible. In this section, we examine the applicability of such approximations.
\subsection{Linearized regime}
In this section, we turn our attention to finding numerical values for the $J$ metric variable. In the Bondi gauge, we have $U_\infty = b_0 = 0$. In such a gauge, Eq.~\eqref{eq:U_infty_constraint} yields
\begin{equation}
C_{30} = -\frac{12 \nu^2 C_{40}}{\ell (\ell + 1) \sqrt{(\ell + 2) (\ell - 1)}} \,.
\end{equation}
In addition, from Eq.~\eqref{eq:E_GW}, we have
\begin{equation}
C_{40} = \frac{(\ell + 2) (\ell - 1)}{6} \left(\frac{\lambda}{2 \pi}\right)^3 \left(\frac{\pi \ell (\ell + 1) \langle\dot{E}_{\text{GW}}\rangle}{2}\right)^{1 / 2} \,.
\end{equation}
We now have the following from Eq.~\eqref{eq:general_J_solution}
\begin{equation}\label{eq:J_ell_ell_bondi_gauge}
J^{[\ell, \ell]} = \sqrt{\frac{32 \pi G \langle\dot{E}_{\text{GW}}\rangle}{c^5}} \left[\frac{i \ell (\ell + 1)}{6} \left(\frac{\lambda}{2 \pi}\right)^2 \sum_{n=4}^{\ell + 2} r^{-(n - 1)} n (n - 3) \Omega_{n, \ell} - \left(\frac{\lambda}{2 \pi r}\right)\right] \,,
\end{equation}
after introducing factors of $G$ and $c$. In the case where $\ell = 2$, Eq.~\eqref{eq:J_ell_ell_bondi_gauge} gives us
\begin{equation}
J^{[2, 2]} = \sqrt{\frac{32 \pi G \langle\dot{E}_{\text{GW}}\rangle_{\ell = 2}}{c^5}} \left[\left(\frac{\lambda}{2 \pi r}\right)^3 - \left(\frac{\lambda}{2 \pi r}\right)\right] \,,
\end{equation}
after making use of $\Omega_{4, 2} = -i \lambda / (8 \pi)$ from Eq.~\eqref{eq:Omega_plus_1}.

We now wish to examine the magnitude of $J^{[2, 2]}$ for the cases of CCSNe and BNS, using the parameter values for $r_i,f$ discussed in Sec.~\ref{sec:astrophysical_applications}. This will be done for the $\ell=2$ case for which there are precise estimates for $\langle\dot{E}_{\text{GW}}\rangle_{\ell = 2}$:
\begin{itemize}
\item \textit{CCSNe:} We take the GW energy to be $10^{39}$J~\cite{Abdikamalov:2020jzn}, and therefore approximate $\langle\dot{E}_{\text{GW}}\rangle_{\ell = 2} \approx 10^{39}$ W. For $f = 100$Hz and $r = 10$km, we have $J^{[2, 2]} \approx 0.181$. On the other hand, for $f=1$kHz and $r = 28$km, we have $J^{[2, 2]} \approx 5.40 \times 10^{-6}$.
\item \textit{BNS:} From~\cite{Bishop:2024oht,Bernuzzi:2015opx}, we take the GW energy values to range from $10^{44}$J to $10^{45}$J, and the timescale in the range $10-50$ms. For $\langle\dot{E}_{\text{GW}}\rangle_{\ell = 2} \approx 10^{47}$W, $f = 1$kHz and $r = 12$km, we have $J^{[2, 2]} \approx 0.98$. On the other hand, for $\langle\dot{E}_{\text{GW}}\rangle_{\ell = 2} = 2\times 10^{45}$W, $f = 2$kHz and $r = 15$km, $J^{[2, 2]} \approx 5.73\times 10^{-3}$.
\end{itemize}
From the form of the angular part of the Bondi-Sachs metric (see, e.g., Eq.~(424) in~\cite{Bishop:2016lgv}), it is clear that nonlinear effects become significant as $|J|\rightarrow 1$. Thus, particularly in the BNS case, there are feasible parameter values for which the linearity assumption would not apply; in these cases, estimates for the GW damping and heating effects may not be reliable.

\subsection{Back-reaction of viscous fluid shell}
In this section, we examine the significance of the back-reaction of viscosity on the fluid flow in the matter shell. We use Cartesian coordinates and consider the metric describing a plane GW along the $z$-direction in the transverse-traceless gauge: $\mathrm{d}s^2 = \mathrm{d}s_\flat^2 + h_{ij} \mathrm{d}x^i \mathrm{d}x^j$, where $h_{11} = -h_{22} =: h_+$, $h_{12} =: h_\times$, and all other components are zero, and where $\mathrm{d}s_\flat^2$ is the Minkowski metric. In the following, we set $h_\times = 0$. The fluid flow is obtained through $V_i = \mathrm{d} X_i / \mathrm{d} t$, where $X_i$ is the proper distance. To obtain the proper distance, we note that, through Taylor expansion, $\sqrt{g_{11}} \approx \left(1 + \tfrac12 h_+\right)$ and $\sqrt{g_{22}} \approx \left(1 - \tfrac12 h_+\right)$. We now have
\begin{equation}
X_1 = x\left(1 + \tfrac12 h_+\right) \,, \hspace{2cm} X_2 = y\left(1 - \tfrac12 h_+\right) \,, \hspace{2cm} X_3 = z \,.
\label{eq:properflow}
\end{equation}
We now set $h_+=A\Re(\mathrm{e}^{i\nu (t-z/c)})$, and differentiate Eq.~\eqref{eq:properflow} to obtain the fluid velocity
\begin{equation}
V_i^{(0)} = 
\left(\tfrac12 x A \nu\Re(i\mathrm{e}^{i\nu (t-z/c)}), -\tfrac12 y A \nu\Re(i \mathrm{e}^{i\nu (t-z/c)}), 0\right) \,,
\end{equation}
where the superscript ${}^{(0)}$ means that the quantity is evaluated when the viscosity $\eta=0$.
We now consider the Navier--Stokes equation:
\begin{equation}\label{eq:Navier_Stokes}
\rho \frac{\partial V_i}{\partial t} = \rho a_i + \eta \nabla^2 V_i \,,
\end{equation}
in the absence of viscosity ($\eta = 0$), and obtain
\begin{equation}\label{eq:a_i_derived}
a_i = \left(-\tfrac12 \nu^2 x A \Re( \mathrm{e}^{i\nu (t-z/c)}), \tfrac12 \nu^2 y A \Re(\mathrm{e}^{i\nu (t-z/c)}), 0\right) \,,
\end{equation}
where $a_i$ is the acceleration caused by the GWs. We now consider $\eta \neq 0$ and write
\begin{equation}
V_i^{(\eta)} = 
\left(\tfrac12 x B \nu\Re(i \mathrm{e}^{i\nu (t-z/c)}), -\tfrac12 y B \nu\Re(i\mathrm{e}^{i\nu (t-z/c)}), 0\right) \,.
\label{eq:Vi_eta} 
\end{equation}
Then substituting Eqs.~\eqref{eq:a_i_derived} and \eqref{eq:Vi_eta} into Eq.~\eqref{eq:Navier_Stokes}, we obtain
\begin{equation}\label{eq:A_eta_to_A0}
B = A + i B\frac{\nu \eta}{\rho c^2} \,.
\end{equation}
Rearranging Eq.~\eqref{eq:A_eta_to_A0} yields
\begin{equation}
B = A\left(1 - \frac{i \nu \eta}{\rho c^2}\right)^{-1} \,.
\end{equation}
It follows that the presence of viscosity will not significantly modify the fluid flow provided that $2 \pi  f \eta / (\rho c^2) \ll 1$.

We now examine the magnitude of $2 \pi  f \eta / (\rho c^2)$ for the cases of CCSNe and BNS, using the parameter values for $f,\eta$ discussed in Sec.~\ref{sec:astrophysical_applications}.
For the density, we use values that are consistent with~\cite{Bishop:2024oht} (see Tables I and III therein):
\begin{itemize}
\item \textit{CCSNe:} Using $\eta = 10^{23}$kg$/$m$/$s and $\rho = 10^{14}$kg$/$m$^3$, we obtain $2 \pi f \eta / (\rho c^2) \approx 6.99 \times 10^{-5}$ for $f=1$kHz, and $2 \pi f \eta / (\rho c^2) \approx 6.99 \times 10^{-6}$ for $f=100$Hz.
%We can also examine this quantity when taking the viscosity to be at the high end of its expected value, which is $\eta = 10^{25}$kg$/$m$/$s~\cite{Bishop:2024oht}. Such a value for the viscosity gives $2 \pi f \eta / (\rho c^2) \approx 3.85 \times 10^{-3}$. We also note that we have used a density value that is on the lower end of the expected range, and on the high end it is approximately $\rho = 10^{16}$kg$/$m$^3$~\cite{Bishop:2024oht}. Using such a value for $\rho$ would result in smaller values for $2 \pi f \eta / (\rho c^2)$.
\item \textit{BNS:} Using $\rho = 10^{16}$kg$/$m$^3$ and $\eta = 10^{24}$kg$/$m$/$s, we obtain $2 \pi f \eta / (\rho c^2) \approx 1.40 \times 10^{-5}$ for $f=2$kHz, and $2 \pi f \eta / (\rho c^2) \approx 6.99 \times 10^{-6}$ for $f=1$kHz.
%and $f = 1500$Hz gives $2 \pi f \eta / (\rho c^2) \approx 1.05 \times 10^{-5}$.
%Once again, this value was obtained using a viscosity value on the low end of the expected range, and on the high end the viscosity is approximately $10^{30}$kg$/$m$/$s~\cite{Bishop:2024oht}. Using such a value for the viscosity gives $2 \pi f \eta / (\rho c^2) \approx 10.5$. While this value is greater than unity, we note that the results presented in the main text for the BNS case were obtained using a value of $\eta = 10^{24}$kg$/$m$/$s.
\end{itemize}
As discussed in~\cite{Bishop:2024oht}, there are estimates in the literature that give larger values for the viscosity $\eta$, up to $10^{30}$kg$/$m$/$s, This would lead to a value of $2 \pi f \eta / (\rho c^2)>1$ and would mean that the back reaction of viscosity would significantly reduce the velocity and thus the shear of the fluid flow.

\section{Computer codes}\label{sec:computer_algebra}
The computer algebra codes used in this work are available as Supplemental Material~\cite{supplemental_material}, and make use of \texttt{sympy}~\cite{10.7717/peerj-cs.103}. We have also made use of \texttt{numpy}~\cite{harris2020array}, and \texttt{matplotlib}~\cite{Hunter:2007} for producing Figs.~\ref{fig:fig_HMNS} and~\ref{fig:fig_CCSNe}. These computer algebra and plotting codes are organized into four packages, as well as an additional \texttt{notebooks} directory which contains the Jupyter notebook: \texttt{beyond\_quadrupolar\_notebook.ipynb}. The Jupyter notebook is used to execute the code contained in the four aforementioned packages. Below, we describe the four packages.

\begin{itemize}
\item\texttt{geometric}:
This package is used for computing the Christoffel symbols as well as the Riemann tensor, Ricci tensor, and Einstein tensor components for a given metric tensor. The package also contains functions that are used for linearizing expressions. In this manuscript, we use this package for the case where the metric is the Bondi--Sachs one given in Eq.~\eqref{eq:Bondi_Sachs_metric}.

\item \texttt{eth\_formalism}:
This package is used for the implementation of eth operations specific to the case where the complex dyad is $q^A = (1, i / \sin\theta)$. In addition, the package contains substitution rules which we use to rewrite certain expressions in terms of eth operations rather than explicit $\theta$ and $\phi$ derivatives. We also use this package for generating $\,_sZ_{\ell, m}$ components.

\item \texttt{eth\_field\_equations}:
This package makes use of the Ricci tensor components generated using the \texttt{geometric} package, and rewrites the expressions in terms of eth operations by making use of the substitution rules contained in the \texttt{eth\_formalism} package. We note that these field equations are obtained for the case where the background fields for $\beta$ and $W$ are arbitrary functions of $r$, and in the present manuscript we have set both of these to zero, which corresponds to considering perturbations around a Minkowski background. As was done in~\cite{Bishop:2019ckc}, we have made use of the following field in our computer algebra
\begin{align}\label{eq:RJ_def}
R_J = (q^A q^B -r^2 J h^{A B})R_{A B}\,,
\end{align}
which coincides with $q^A q^B R_{AB}$ when the background is Minkowski. In addition to the linearized field equations, the aforementioned substitution rules are also implemented to obtain the matter Eqs.~\eqref{eq:B_11}--\eqref{eq:Theta}.

This package is also used for inserting the ansatz for the metric and velocity field components into a given expression, and reducing the field equations to differential equations for the radial parts. The final $\ell$-mode expressions for the metric variables and shear tensor components given in Eqs.~\eqref{eq:general_U_solution}--\eqref{eq:U_infty_constraint}, and~\eqref{eq:sigma_11_ell}--\eqref{eq:sigma_U_ell}, respectively, are also included. The evaluation of these arbitrary $\ell$-mode expressions is done by making use of the following expression for the $\Omega_{n, \ell}$ coefficients when $n > 1$:
\begin{align}\label{eq:Omega_def}
\Omega_{n, \ell} = \frac{3 i \nu}{\ell (\ell + 1) (n - 1)} \prod_{m = 3}^n \frac{(m - 1) (\ell + m - 2) (\ell - m + 3)}{2 m (m - 2) i \nu} \,, \hspace{1.5cm} n > 1 \,.
\end{align}
This evaluation is used to generate the expressions for the metric and matter variables given in Appendix~\ref{sec:specific_ell_metric}. It is also used to perform fixed-$\ell$ consistency checks of the field equations.
\item \texttt{fixed\_ell}:
This package is used to generate the rescaled GW magnitude and temperature increase for a fixed value of $\ell$, as well as the damping factors. It is also used for the numerical evaluation of the damping factors in the cases where $\ell = 2$, $3$, or $4$, and producing the plots given in Figs.~\ref{fig:fig_HMNS}, and~\ref{fig:fig_CCSNe}.
\end{itemize}
For further details regarding the computer codes used in the present work, we direct the interested reader to the \texttt{beyond\_quadrupolar\_notebook.ipynb} Jupyter notebook.
\end{document}